\documentstyle[12pt,aasms]{article}

\def\lsim{\lower.5ex\hbox{$\; \buildrel < \over \sim \;$}}
\def\gsim{\lower.5ex\hbox{$\; \buildrel > \over \sim \;$}} 

\begin{document}

\title{Grand Unification of Solutions of Accretion and Winds
around Black Holes and Neutron Stars}

\author{Sandip K.\ Chakrabarti$^1$} 
\affil{Goddard Space Flight Center, Greenbelt MD, 20771\\
and 
Tata Institute of Fundamental Research, Homi Bhabha Road, Bombay, 400005$^2$\\
e-mail: I: chakraba@tifrvax.tifr.res.in}

\noindent: Submitted June 7th, 1995; Appearing in ApJ on June 20th, 1996.\\

\begin{abstract}

We provide the complete set of global solutions of viscous transonic 
flows (VTFs) around black holes and neutron stars. These solutions describe the
optically thick and optically thin flows from the
horizon of the black hole or from the neutron star surface to
the location  where the flow joins with a Keplerian disk.  We study the nature
of the multiple sonic points as functions of advection, rotation, viscosity,
heating and cooling. Stable shock waves, which 
join two transonic solutions, are found to be present
in a large region of the parameter space. We classify the
solutions in terms of whether or not the flow can have a 
standing shock wave.
We find no new topology of solutions other than what are
observed in our previous studies of isothermal VTFs.
We particularly stress the importance of the boundary conditions
and argue that we have the most complete solution of
accretion and winds around black holes and neutron stars.

\end{abstract}
 

\noindent $^1$ NRC Senior Research Associate at GSFC\\
\noindent $^2$ Permanent Address\\

\newpage
\section{INTRODUCTION}
 
Standard accretion disk models of Shakura \& Sunyaev (1973; hereafter SS73)
and Novikov \& Thorne (1973) have been very useful in interpretation
of observations in binary systems and active galaxies (e.g. Pringle 1981;
Shapiro \& Teukolsky, 1984; Frank et al., 1992). 
The description of physical quantities in these
models are expressed analytically and they could be used 
directly. However, these models do not treat the pressure
and advection terms correctly, since the disk is terminated at the
marginally stable orbit (three Schwarzschild radii for a
non-rotating black hole) and no attempt was made to satisfy the
inner boundary condition on the horizon. A second problem
arose, when it was pointed out (Lightman \& Eardley,
1974) that the inner regions of these disks are viscously
and thermally unstable. Observationally, there are overwhelming
evidences that the disks are {\it not} entirely Keplerian (see,
Chakrabarti 1993, 1994, 1995, 1996, hereafter C93; C94; C95;
and C96 respectively).
The soft and high states of the galactic and extragalactic black hole
candidates (Tanaka et al., 1989; Ebisawa et al., 1994)
are very poorly understood, and it has been suggested very recently 
(Chakrabarti \& Titarchuk 1995, hereafter CT95) that this change of
states could be attributed to the presence of the sub-Keplerian 
components which may include shock waves. The general agreements of
the prediction of CT95 with observations strongly suggest the
reality of sub-Keplerian advective flow models.

Paczy\'nski and his collaborators (Paczy\'nski \& Bisnovatyi-Kogan,
1980; Paczy\'nski \& Muchotrzeb, 1982) have attempted to include
advection and pressure effects in the so-called transonic accretion
disks, although no systematic study of global solutions were performed.
Global solutions of the so-called `thick accretion disks' were possible to
obtain only when the advection term is dropped (e.g., Paczy\'nski \&
Wiita, 1980). In these accretion disks, the flow is assumed to have 
practically constant angular momentum. Some exact solutions
of fully general relativistic thick disks are discussed in 
Chakrabarti (1985; hereafter C85). 

Early attempts to find global solutions of viscous transonic flow (VTF)
equations (Muchotrzeb, 1983; Matsumoto et al. 1984) concentrated 
much on the nature of the inner sonic point of these
flows which is located around the marginally stable orbit.
In the case of inviscid adiabatic flow, an example of global 
solution was provided by Fukue (1987) who performed a
study of shocks similar to that in solar winds and galactic jets
(e.g., Ferrari et al., 1985) and found evidence of shock transition
as well. In the so-called `slim-disk' model of
Abramowicz et al. (1988), it was tried to show from {\it local 
solutions} that the instabilities at the inner edge could be removed
by addition of the advection term (see, a similar trial by Taam
\& Fryxall, 1985; Chakrabarti, Jin  \& Arnett, 1987,
 where thermonuclear reaction in the disk
was used to eliminate the instability). The global solution 
of Abramowicz et al.  was not
completely satisfactory to the present author, since the angular
momentum, instead of joining to Keplerian, {\it deviated away} from it
close to the outer edge (see, Fig. 3 of Abramowicz et al. 1988).
First satisfactory global solution of these equations in the
optically thin or thick limit which include
advection, viscosity, heating and cooling in the limit of isothermality
condition was obtained by Chakrabarti 
(1990a, hereafter C90a; 1990b, hereafter C90b) where  disk
models of (single) temperature ($\gsim 1.e+11$K) which 
become Keplerian far away were considered. Recent self-consistent
Comptonization work of CT95 shows that in the presence of soft-photon source
from Keplerian component, the protons can be isothermal (see, Fig. 2 of CT95)
in some range of accretion rates ($\sim 0.3-0.5$ times the Eddington rate)
and therefore isothermality condition of C90a,b may be more realistic 
than thought before.

In an earlier work (Chakrabarti 1989, hereafter
C89), we have presented the complete classification of global solutions of an
{\it inviscid}, polytropic transonic flow (see, Fig. 4 of C89) which showed 
that in some region of the parameter space, the flow will have multiple sonic
points (e.g., Liang \& Thomson, 1980). We also found that within this 
region, there is a sub-class of solutions where Rankine-Hugoniot
shock conditions are satisfied and shock waves are formed due to the
centrifugal barrier (centrifugally supported shocks). These shock
solutions are perfectly transonic. Matter inflowing
into a black hole crosses a sonic region three times, twice (continuously)
at the outer and the inner sonic points, and once (discontinuously)
at the shock location. Four locations, namely, $x_{si},\  (i=1..4)$
were identified where these shocks could formally be located, but it was
pointed out that only $x_{s2}$ and $x_{s3}$ were important
for accretion on black holes since 
the flow has to be supersonic on a black hole horizon and $x_{s1}$ could
also be important for a neutron star accretion while $x_{s4}$ was a purely 
formal shock location. In Chakrabarti 90a,and 90b, 
viscosity was also added and complete global solutions in isothermal VTFs
with and without shocks, were found. In the language of Shakura-Sunyaev 
(SS73) viscosity parameter $\alpha$, it was shown that if
viscosity parameter is less than some critical value $\alpha_{cr}$,
the incoming flow may either have a continuous 
solution passing through outer sonic point, or, it can have standing shock 
waves at $x_{s3}$ or $x_{s2}$ (following notations of C89 or C90a,b)
if the flow allows such a solution in accretion.
For $\alpha >\alpha_{cr}$, a standing shock wave at $x_{s2}$
persisted, but the flow now had two continuous solutions --- one
passed through the inner sonic point, and the other through the
outer sonic point. Later analytical and numerical works
(Chakrabarti \& Molteni, 1993, 1995; hereafter CM93 
and CM95 respectively, Nobuta \& Hanawa 1994, Nakayama 1992) 
showed that $x_{s3}$ is stable, and that for $\alpha>\alpha_{cr}$ the 
continuous solution passing through the inner sonic point is chosen. 
We noted that $\alpha_{cr}$ ($\sim 0.015$ for the isothermal case considered)
was a function of the model parameters, such as the disk temperature, 
sonic point location and angular momentum on the horizon.  Most importantly,
these solutions show that they join with the Keplerian disks at some 
distance, depending upon viscosity and angular momentum (C90a, CM95). 
This discussion of critical $\alpha$ is valid
when the inner sonic point and angular momentum of the
flow at the inner edge is kept fixed (see below).

Extensive numerical simulations of quasi-spherical, inviscid,
adiabatic accretion flows (Molteni, Lanzafame \& Chakrabarti 1994; hereafter
MLC94), show that shocks form very close to the location
where vertically averaged model of adiabatic flows predict 
them (C89). The flow advected its entire energy to the black hole
and the entropy generated at the shock is also totally advected 
allowing the flow to pass through the inner sonic point. 
It was also found, exactly as predicted in C89,
that flows with positive energy and higher entropy form strong,
supersonic winds. In presence of viscosity also,
very little energy radiates away (e.g., Fig. 8 of C90a).
Having satisfied ourselves of the stability of these solutions (CM93, 
MLC94, CM95), we proposed a unified scheme of accretion disks 
(C93, C94; C96, and CM95) which combines
the physics of formation of sub-Keplerian disks with and without shock
waves depending on viscosity parameters and angular momentum at the
inner edge. We always considered only the stable branch of VTF
and our solutions remained equally valid for black hole and neutron 
star accretions as long as appropriate inner boundary conditions are employed.
Importance of these findings are currently being reconsidered in the so-called
`newly discovered advection dominated model' (Narayan \& Yi, 1994; 
see Narayan, 1996 and references therein).

In this paper, we make a comprehensive study of the global solutions of the
VTF equations applicable to black hole and neutron star
accretion. We remove the restriction of isothermality condition
imposed in C90a and 90b, and made explicit use of the energy equation. 
We classify the solutions 
according to whether or not an accretion flow can have shock waves. 
We include the effects of advection, rotation, viscosity, heating and cooling
as before. 
We discover the existence of two critical viscosity parameters:
$\alpha_{c1} (x_{in}, l_{in})$ and $\alpha_{c2} (x_{in}, l_{in})$ 
which control the nature of the inner regions of the disk (Here,
$x_{in}$ and $l_{in}$ denote the inner sonic point and the 
angular momentum of the flow on the horizon or star surface respectively.)
Out of these two, $\alpha_{c2}$ has the same meaning (shock/no-shock)
as $\alpha_{cr}$ in isothermal case (C90a)
whereas $\alpha_{c1}$ (also present in isothermal case, but we did not 
explore it before) determines whether the flow would be in the accretion
shock regime in the first place. We assume 
the standard viscosity type prescription (SS73), but the
viscous stress is assumed to be proportional to the thermal pressure
(standard assumption of SS73) or the total (thermal plus ram)
pressure (CM95). The latter is useful when advection (radial
velocity) is important. We also examine the effects of the polytropic
index of the flow on the sonic point behavior and note that 
typically, for $\gamma \lsim 1.5$ there are multiple
sonic points (see, Fig 3.1 of C90b).
For $\gamma \gsim 1.5$, or generally for higher viscosity
or lower cooling efficiency the outer saddle
type sonic points are absent and therefore shocks could only form
if the flow is already supersonic (such as coming from some stellar winds).
In all these cases, even with general heating and cooling, we do not discover
any new topologies other than what are already discovered in C90a and C90b. 
We argue in Section 5 that there should not be any new topologies
either. We therefore believe that the present result
contains the most complete solutions to date which one may have around a 
black hole or a neutron star. We do not consider accretion through nodal
points (Matsumoto et al., 1984) as the stability properties
of these solutions are uncertain.

Throughout the paper, we give importance to two fundamental issues
related to a black hole and neutron star accretion: the nature of the 
sonic points, and the typical distances ($x_{Kep}$) at which the disk 
may join a Keplerian disk. Understanding of the nature of the sonic point is 
important, since matter accreting on a black hole
must pass through it (C90b). Similarly, knowledge
of how $x_{Kep}$ depends on viscosity is very crucial because of 
the possible role it may have on the observed high energy
phenomena, such as novae outbursts and soft and high states of galactic 
black hole candidates. The spectra would be a mixture of the 
emission from Keplerian and non-Keplerian components (CT95) and we need to
know at what distance the deviation from Keplerian distribution becomes
important. These non-Keplerian flows have been exactly solved using
the sonic point analysis and their properties studied
extensively in the past few years using restricted equation of states 
(C89, C90a, C90b). In the present paper, we only extend these studies
to include more general heating and cooling processes.
Non-axisymmetric, non-Keplerian, vertically averaged flows which are 
more difficult to deal with have been solved using self-similarity assumption
(Chakrabarti 1990c; see also Spruit, 1987 who used self-similarity
for a disk of constant height or conical disk without vertical averaging.)
with polytropic equation of state. These studies in the present 
context of more general heating and cooling will be presented in future.

In our analysis of black hole and neutron star accretion, we use
the Paczy\'nski-Wiita (1980) potential. This potential is known
to mimic the geometry around a Schwarzschild black hole quite
satisfactorily and is widely used in the astrophysical community.
One major misgivings of our present work may be that we do not use general
relativity (GR).
Our prior experience of solving inviscid disks in Kerr geometry
(Chakrabarti, 1990d) indicates that no new topological properties emerge 
when full general relativity is used. Even in magnetohydrodynamical
studies (Takahashi et al., 1990; Englemaier, 1993) no new topology
emerges other than what is observed with pseudo-Newtonian potential
(Chakrabarti, 1990e). The generalized equations in Kerr geometry using 
the prescription of Novikov \& Thorne (1973), but with
conserved angular momentum $l=-u_\phi/u_t$, do not yield
any new topologies either (Chakrabarti, 1996b). Only 
quantitative change is the possible reversal
of shear stress just outside the horizon (Anderson \& Lemos, 1988).
In GR one could apply the inner boundary condition (`lock-in' of 
the flow with the horizon) rigorously than what we could do with 
pseudo-Potential. For instance, the definition of
angular velocity of matter is related to the angular momentum
($\Omega$) by (e.g., C85):
$$
\Omega = \frac{l}{\lambda^2}
$$
where,
$$
\lambda^2=-\frac{u_\phi u^t}{u_t u^\phi} = \frac{x^2 sin^2\theta}
{1-\frac{1}{x}} 
$$
(The second equality is valid for Schwarzschild geometry. Here,
$u^\mu$s are the four velocity components and $l$ is the specific
angular momentum.) 
Thus, by definition, independent of how much angular momentum is
carried in by the flow, the flow would have `zero' angular velocity
on the horizon ($x=1$). However, when using a pseudo-potential, this
condition is not met: $\Omega = l_{in}/x^2$ is a Newtonian definition
($l_{in}$ being the angular momentum of the inflow on the horizon)
and it does not vanish at $x=1$, the horizon (here distance $x$
is measured in units of $x_g=2GM_{BH}/c^2$). Usually 
this poses no threat. The potential energy is infinite at $x=1$
in this potential, hence, the rotational energy term $\Omega^2 x^2/2$
is always insignificant at $x=1$.
The same consideration of `locking-in' condition
for a Kerr black hole implies that the rotational velocity of the flow
`matches' with that of the black hole at the horizon (e.g., Novikov
\& Thorne, 1973). A pseudo-potential
has been constructed with this consideration (Chakrabarti
\& Khanna, 1992) which is valid for small Kerr parameter `a' only. 
For a neutron star solution, the situation is simpler: one just
has to choose the final subsonic branch such that on the
star surface, $x=R_*$, $\Omega_* R_*^2 = l_{in}$ condition is satisfied.
In the numerical simulations (CM93, MLC94, CM95) the inner boundary
condition is achieved by putting an `absorption' boundary condition
at $x\sim 1$.

The plan of the paper is the following: in the next section, we present 
model equations, description of each terms and show why
such flows would have multiple sonic points. In \S 3,
we present analytical solutions for flow parameters
at the sonic points using both the viscosity prescriptions. 
In \S 4, we present results depicting extensively how 
the solution topologies depend on the disk parameters. 
In \S 5, we briefly argue about the completeness of our solutions.
Finally, in \S 6, we make concluding remarks.

\section{MODEL EQUATIONS}

We assume units of length, velocity and time to be $x_g=2GM_{BH}/c^2$, $c$ 
and $2GM_{BH}/c^3$ respectively. We assume the flow to be vertically 
averaged. 
Comparison of numerical simulations  of thick accretion (MLC94) with 
analytical results (C89) indicate that vertically averaged treatments 
are basically adequate. The equations of motion
which we employ (CT95, C96) are similar to, but not exactly same as those
used by Paczy\'nski \& Bisnovatyi-Kogan, 1981; Matsumoto et al. 1984;
Abramowicz et al. 1988; C90a,b; Narayan \& Yi, 1994. We use,

\noindent (a) The radial momentum equation:

$$
\vartheta \frac{d\vartheta}{dx} +\frac{1}{\rho}\frac{dP}{dx} 
+\frac {l_{Kep}^2-l^2}{x^3}=0,
\eqno{(1a)}
$$

\noindent (b) The continuity equation:

$$
\frac{d}{dx} (\Sigma x \vartheta) =0 ,
\eqno{(1b)}
$$

\noindent (c) The azimuthal momentum equation:

$$
\vartheta\frac{d l(x)}{dx} -\frac {1}{\Sigma x}\frac{d}{dx} 
(x^2 W_{x\phi}) =0 ,
\eqno{(1c)}
$$

\noindent (d) The entropy equation:

$$
\Sigma \vartheta T \frac{ds}{dx} = \frac{h(x) \vartheta}{\Gamma_3 - 1} (
\frac{dP}{dx} - \Gamma_1 \frac {P}{\rho}\frac{d\rho}{dx})=
Q^+ - Q^- = \alpha q^+-g(x, {\dot M}) q^{+}
=f(\alpha, \ x,\ {\dot M}) Q^{+},
\eqno{(1d)}
$$

where, 
$$
\Gamma_3=1+\frac{\Gamma_1-\beta}{4-3\beta},
$$
$$
\Gamma_1=\beta + \frac{(4-3\beta)^2 (\gamma -1 )}{\beta + 12 (\gamma -1)(1-\beta)}
$$
and $\beta (x) $ is the ratio of gas pressure to total pressure,
$$
\beta= \frac{\frac{\rho k T}{\mu m_p}}{\frac{1}{3} {\bar a} T^4 + \frac{\rho k T}
{\mu m_p}} .
$$
Here, ${\bar a}$ is the Stefan constant, $k$ is the Boltzman constant, $m_p$
is the mass of the proton and $\mu$ is the mean molecular weight. Note that
for a radiation dominated flow, $\beta \sim 0$, 
and $\Gamma_1=4/3=\Gamma_3$ and for a gas pressure dominated flow, 
$\beta \sim 1$, and $\Gamma_1=\gamma=\Gamma_3$.
Using the above definitions, eqn. (1d) becomes,
$$
\frac{4-3\beta}{\Gamma_1-\beta} [\frac{1}{T}\frac{dT}{dx} 
-\frac{1}{\beta}\frac{ d \beta}{dx} - 
\frac{\Gamma_1 - 1}{\rho}\frac{d\rho}{dx} ] = f(\alpha, x, {\dot M}) Q^+.
\eqno{(1e)}
$$

In this paper, we shall concentrate on solutions with constant $\beta$. Actually
we study in detail only the special cases, $\beta=0$ and $\beta=1$, so we shall liberally use $\Gamma_1=\gamma=\Gamma_3$. Similarly,
we shall consider the case for $f(\alpha, x, {\dot M})$ = constant, though
as is clear, $f\sim 0$ in the Keplerian disk region and probably closer
to or less than $1$ near the black hole depending on the efficiency of
cooling (governed by ${\dot M}$, for instance). If the cooling process
is `super-efficient', namely, when the flow cools faster than it is heated,
$f$ could be negative as well. Two examples of global solutions
with such a possibility, one with 
bremsstrahlung cooling in weak viscosity limit (Molteni, Sponholz,
\& Chakrabarti, 1996; hereafter MSC96) and the other with 
Comptonization (CT95) have been recently discussed. Results of the
general equation (1e) will be presented in near future.
We use Paczy\'nski-Wiita (1980) potential to describe the
black hole geometry. Thus, $l_{Kep}$, the Keplerian angular 
momentum is given by, $l_{Kep}^2=x^3/2(x-1)^2$.
Here, $W_{x\phi}$ is the vertically integrated viscous stress,
$h(x)$ is the half-thickness of the disk at radial
distance $x$ obtained from vertical equilibrium assumption
(C89), $l(x)$ is the specific angular momentum,
$\vartheta$ is the radial velocity, $s$ is the entropy density
of the flow, $Q^+$ and $Q^-$ are the heat gained and lost by the flow,
and ${\dot M}$ is the mass accretion rate.
The constant $\alpha$ above is the Shakura-Sunyaev (SS73)
viscosity parameter which defines the viscous stress as 
$W_{x\phi}=-\alpha W=-\alpha_{P} W$, where $W$ is the integrated
pressure $P$. (We shall refer to this $W_{x\phi}$ as the ``P-stress'' in 
future.) As noted earlier (CM95), instead of having the 
stress proportional to the thermal pressure $W$ as in SS73
it is probably more appropriate to use 
$W_{x\phi} =-\alpha \Pi=-\alpha_{\Pi} \Pi$ while studying flows with 
significant radial velocity, since, especially,
the total pressure (or, the momentum flux) $\Pi=W+\Sigma \vartheta^2$ 
is continuous across the shock and such a $W_{x\phi}$ keeps
the angular momentum across of the shock to be continuous as well.
(We shall refer to this $W_{x\phi}$ as the ``$\Pi$-stress" in future.)
Except for eq. 1d, other equations are the same as used in our 
previous studies (C89 and C90ab). The term $g(x, {\dot M}) \leq \alpha  $ is a
dimensionless proportionality constant, which will be termed as the
cooling parameter. When $g \rightarrow \alpha$, the flow is efficiently
cooled, but when $g\rightarrow 0$, the flow is heating dominated
and most {\it inefficiently} cooled. In C89, eq. 1d was replaced
by the adiabatic equation of state
$P=K\rho^\gamma$ with entropy constant $K$ different in pre-shock
and post-shock flows, and in C90a, eqn. 1d was replaced by the isothermal
equation of state $W=K^2\Sigma$ ($K$ being the sound speed of the gas and 
$\Sigma$ being the integrated density of matter). In the present 
paper, for simplicity, we assume $f(\alpha, x, {\dot M})=(\alpha-g)/\alpha$
=const. We have verified that our conclusions do not change when more general 
cooling laws are used instead. Details will be presented in future.

Though in computing angular momentum distribution, we talked about 
using the shear stress  of the first (1) form, $W_{x\phi} (1)=-\alpha W$,
we wish to note that  a second (2) form, such as, $W_{x\phi} 
(2) =\eta x d\Omega/dx$ (e.g., C90a,b) is also used in the literature. 
However, the latter choice requires an extra boundary condition,
which, in the context of Paczy\'nski-Wiita (1980) potential is
difficult to implement, since on the horizon, various physical quantities
become singular. Thus, in some example of C90a,b, we  chose $d\Omega/dx$ at the
sonic point assuming angular momentum remains almost constant 
between the sonic point and the horizon. While computing the heating term,
$$
Q^+=W_{x\phi}^2/\eta
$$
one could use either $W_{x\phi} (1)$, or , $W_{x\phi} (2)$ or a combination
of both! If only $W_{x\phi} (1)$ is used, no information of `actual shear' is
present in the heating term. If only $W_{x\phi} (2)$ is used 
the equations become difficult to solve, as the sonic point condition
does not remain algebraic any more. (This was not a problem in C90a and C90b;
as the heating equation was replaced by isothermality condition.)
In the present paper, we have chosen to
use the combination of both forms in order that we may be able to
do sonic point analysis with {\it local} quantities (Flammang, 1982) at the
same time retaining the memory of $d\Omega/dx$ of the flow.
We call this prescription as the MIxed Shear Stress (MISStress) prescription.
Roughly speaking, this method is almost equivalent to replacing
one factor of $d\Omega/dx$ by $d\Omega_{Kep}/dx$ while
other factor of $d\Omega/dx$ is kept in tact.
Satisfactory preliminary results with this are already reported in 
the Appendix of CT95  and C96. We have verified that the results with only
$W_{x\phi}(1)$ are very similar.

In order to understand the origin of multiple sonic points, we consider the
property of constant energy surfaces as in the phase-space analysis
in classical mechanics. Integrating eq. (1a) for an isothermal flow
($\gamma=1$), and ignoring resulting slowly varying logarithmic thermal energy
term, we get the specific energy of the flow to be,
$$
E=\frac{1}{2} \vartheta^2 + \frac{1}{2} \frac{l^2}{x^2} -\frac{1}{2(x-1)} .
$$
Note that since the potential energy term (third term)
is dominant compared to the rotational energy term (second term)
both at a large distance $x\rightarrow \infty$ as well as
close to the horizon $x \rightarrow 1$, the behavior of
constant energy contours in the phase space ($\vartheta - x$ plane,
or, equivalently, in $M-x$ plane for isothermal flows)
is {\it hyperabolic} (because of the negative sign in front of $(x-1)^{-1}$
term) and as a result, saddle type
sonic point is formed. In the intermediate distance, $l \sim x$, the
rotational term is dominant and the constant energy contours 
are elliptical (because of the positive sign in front of $l^2/x^2$ term)
and center type sonic point is formed (C90b).
Of course, whether or not all these three sonic points will be present
depends on the angular momentum, polytropic index $\gamma$,
viscosity, heating and cooling effects. 

From the continuity equation (1b), we obtain the mass accretion rate
to be given by,
$$
{\dot M}=2\pi\rho h(x) \vartheta
\eqno{(2)}
$$
and from the azimuthal momentum equation,
$$
l-l_{in}=\frac{\alpha_{P}}{\gamma} \frac{x}{\vartheta} a^2
\eqno{(3a)}
$$
(see C90a, C90b) using P-stress prescription: $W_{x\phi}=-\alpha_P W$, or,
$$
l-l_{in}={\alpha_{\Pi}} \frac{x}{\vartheta} a^2 [\frac{2}
{3\gamma -1} +  M^2 ]
\eqno{(3b)}
$$
for the $\Pi$-stress prescription (CM95) $W_{x\phi}=-\alpha_{\Pi} 
(W+\Sigma\vartheta^2)$. Here $M=\vartheta/a$ is the Mach number
of the flow, $a$ being the sound speed (defined by $a^2=\gamma P/\rho$).
The thickness of the disk is $h(x)\sim a x^{1/2}(x-1)$.
The integration constant $l_{in}$ represents the
angular momentum at $x=1$ which we refer to as the specific angular 
momentum at the inner edge of the flow, namely on the horizon.
For a neutron star accretion the inner boundary condition has to be
$l_{in}=\Omega_* R_*^2$ at $x=R_*$ as discussed earlier.
Since in the regime of the pseudo-Newtonian potential, $v \rightarrow 
\infty$ and $a$=finite on the horizon, this identification is justified.
The derivation of the squared bracketed terms of eq. (3b) requires an 
understanding of the vertically averaged quantities (C89, Matsumoto 
et al. 1984),
$$
\Sigma = \int_{-h/2}^{h/2} \rho dz =  \rho_e I_n h \ \ \ \ 
W=\int_{-h/2}^{h/2} p_e I_{n+1} h=p_e I_{n+1} h
\eqno{(4)}
$$
where, $\rho_e$, and $p_e$ are the equatorial quantities. In terms of
$n=1/{\gamma-1}$, $I_n$ is given by,
$$
I_n=\frac{(2^n n\!)^2}{(2n+1)}.
\eqno{(5)}
$$

Below, we present results using both the P-stress and $\Pi$-stress
prescriptions. Though generally results are similar, angular momentum
is found to be continuous across the shock when we use $\Pi$-stress,
whereas it is discontinuous when we use P-stress. Thus,
the shock becomes purely compressible type when ram pressure is included in
the stress, whereas it becomes a mixture of shear and compressible types when
only thermal pressure is included (CM95).
The other difference is that since there are two terms
in $\Pi$-stress which come from the thermal and ram pressures, $\alpha_{P}
\sim 2 \alpha_{\Pi}$. We shall discuss both types of
flows here for completeness, though we shall study
examples of purely compressible shocks for simplicity using $\Pi$-stress. 

\section{SONIC POINT ANALYSIS}

We solve equations 1(a)-1(e) 
using a sonic point analysis as before (C89, C90a,b).

\subsection{Results Using ``P-Stress"}

This analysis is done with $W_{x\phi}=-\alpha_P W$. This is used
in computing the angular momentum distribution (3a). However, 
in computing $Q^+$, we have used MISStress prescription.
Here, one has
$$
Q^+=\alpha P h(x) x \frac{d\Omega}{dx} .
$$
Hence eq. (1e) becomes,
$$
\frac{2n}{a}\frac{da}{dx} - \frac{1}{\rho}\frac{d\rho}{dx}={\alpha_P 
f(\alpha, \ x,\ {\dot M})}{\vartheta} \frac{d\Omega}{dx}.
$$
The function $f$ will be close to zero in the Keplerian disk at the
outer edge due to efficient cooling process
and $f<1$ close to the black hole (it could even be negative for super-cooled
system, especially for weakly viscous case, either for optically thin flows
with bremsstrahlung emission, MSC96, or, 
for optically slim flows with Comptonization process,  CT95). 
But we choose it to be a constant
for simplicity, and also for clarity we use $\alpha$ in stead of 
$\alpha_P$ in the rest of this section. After some algebra we find,
$$
\frac{d\vartheta}{dx}= \frac{N}{D},
\eqno{(6)}
$$
where the numerator is,
$$
N=\left[\frac{1}{2(x-1)^2}-\frac{l^2}{x^3}-\frac{a^2}{\gamma}\frac{5x-3}
{2x(x-1)}\right]\left[\frac{(2n+1)\gamma}{a^2}-\frac{2\alpha^2 f}
{\vartheta^2}\right] -\left[\frac{5x-3}{2x(x-1)}+\frac{2l\alpha f}
{\vartheta x^2} - \frac{\alpha^2 f a^2}{\gamma x \vartheta^2} \right]
\eqno{(7a)}
$$
and the denominator is,
$$
D=\left[\frac{(2n+1)\gamma}{a^2} - \frac{2 \alpha^2 f}{\vartheta^2} \right ]
\left[ \frac{a^2}{\gamma \vartheta}-\vartheta \right] +
\left[ \frac{1}{\vartheta} + \frac{\alpha^2 f a^2}{\gamma \vartheta^3}
\right ] .
\eqno{(7b)}
$$
Here, $n=(\gamma-1)^{-1}$, $\gamma$ being the polytropic index.
At the sonic point, the numerator and the denominator both vanish. 
From $D=0$, one obtains the Mach Number $M_c (x_c)$,
$$
M_c^2(x_c)= \frac{[\alpha^2 f +n+1+\{ \alpha^2 f (\alpha^2 f + 1) +
(n+1)^2 \}^{1/2}]}{[\gamma (2n+1)]} \approx \frac{2n}{2n+1} 
{\rm \ \ \ for \ \alpha <<1}
\eqno{(8a)}
$$
For $N=0$, one obtains an exact expression for the velocity of sound $a$,
$$
a_c (x_c)=\frac{2{\cal A}}{({\cal B}^2 - 4 {\cal A}{\cal C})^{1/2}-{\cal B}}
\eqno{(8b)}
$$
where,
$$
{\cal A}= \gamma (2n+1-\frac{2\alpha^2f}{\gamma M_c^2})
[\frac{1}{2(x-1)^2}-\frac{l_{in}^2}{x^3}]
$$
$$
{\cal B}=-\frac{2 l_{in} \alpha } {x^2 M_c}
(2n+1+f-\frac{2\alpha^2f}{\gamma M_c^2})
$$
and
$$
{\cal C}= -(2n+2-\frac{2\alpha^2 f}{\gamma M_c^2})\frac{5x-3}{2x(x-1)}
-\frac{\alpha^2}{\gamma M_c^2 x}(2n+1+f-\frac{2\alpha^2 f}{\gamma M_c^2})
$$
For $\alpha=0$, eq. (8a) goes over to
$M_c^2(x_c)=2n/(2n+1)$ as in C89 and over to $M(x_c)=1$ for $\gamma=1$
(isothermal flow) as in C90a. Similar limit is obtained for
$a_c(x_c)$ as well. These conditions allow us to obtain solutions
with one {\it less} parameter, since two extra vanishing conditions
of numerator and denominator provide two equations while
only one extra unknown ($x_c$) is introduced. The number of parameters
required to study shocks does not change (Abramowicz \& Chakrabarti, 1990;
C89, C90b). Sonic points occur only where the sound speed is real and positive
(Liang \& Thomson, 1980; Abramowicz \& Zurek, 1981) and the saddle type 
sonic points (which are vital to a global solution, see C90b)
occur when $d\vartheta/dx|_c $ is real and of two different signs
(Thompson \& Stewart, 1985; Ferrari et al. 1985; C90b).
It is easy to see that above constraints can allow a maximum of two saddle
type sonic points (vital to a shock formation) when $\gamma \lsim 1.5$.
For higher $\gamma$, only inner saddle type sonic point may form,
depending on viscosity and accretion rates (responsible for cooling
parameter $g$) unless the flow is of constant height, in which case both
sonic points will form even for $\gamma=5/3$ (MSC96). 

In order to study shock waves around a black hole, 
it is crucial to know if the flow has more than one saddle type sonic point.
In a neutron star accretion, one saddle type point is sufficient.
This is due to the fundamental difference in the
inner boundary conditions. In a black hole accretion, the flow
first passes through the outer sonic point, and then, if it passes
through a shock, the flow becomes subsonic. It has to pass through
another sonic point to satisfy the `supersonic' boundary condition on the
horizon with the radial velocity equal to the velocity of light 
and the rotational velocity locked-in with the rotational velocity
of the horizon.
At a neutron star boundary, the flow is subsonic, and thus
after a shock the flow need not pass through a sonic point. If it
does, however, it has to have another shock to satisfy the inner 
boundary condition and the corrotation condition with the star surface. 
We shall present exact solutions of these kinds
in later sections. Not only the flow should have two saddle type points, 
the entropy at the inner sonic point should be higher compared to the entropy
at the outer sonic point and the
energy at the inner sonic point must be smaller or equal to the
energy at the outer sonic point (C89). For a wind flow,
these considerations are exactly the opposite. Thus, one must know the 
nature of the energy and entropy densities at the sonic points.

Fig. 1a shows entropy $s(x_c) \propto a(x_c)^{2n}/\rho(x_c)$ {\it vs.}
specific energy $E(x_c)=0.5 \vartheta^2+
(\gamma-1)^{-1} a(x_c)^2 + 0.5\  l(x_c)^2/x_c^2 - 0.5\ (x_c-1)^{-1}$ plots 
at the {\it sonic points} for $\gamma=4/3$ and
$\alpha_P=0, \ 0.4 \ 0.8$. The arrows on the curves in the direction
of increasing sonic point location $x_c$. Two side-by-side curves
are for $f=0$ (cooling dominated) and $f=1$ (heating dominated) 
respectively (as marked on the curves). For $\alpha_P=0$, these
two curves coincide. Solid, dashed and dotted
regions of the curves are the saddle, nodal and spiral
(circle type for $\alpha_P=0$) type sonic points respectively
(C89, C90a,C90b). The branches $AMB$ and $CMD$
(while referring similar branches of curves with
non-zero viscosity and heating parameter $f$, we shall 
use same notations such as $AM$, $MB$ etc., though we did
not mark them on this plot for clarity) are the results on
the inner ($x_{in}$) and the outer ($x_{out}$) sonic points with
$M$ being the point where an inviscid flow can pass through
the inner and outer sonic points simultaneously. The general ``swallow
tail singularity'' as seen in these figures was noticed by 
Lu (1985) though importance of
having different entropy at different sonic points was not noted
as a result it was thought that a flow with the same outer
boundary condition may have multiple solutions and bi-periodicities 
(e.g. Abramowicz \& Zurek, 1981; Lu, 1985). 
Parameters around the point $M$ are 
important to study shock waves in the flow (C89, C90a, C90b). Note that 
as viscosity is increased, more and more region containing
saddle type sonic points become spiral and nodal type, particularly, the 
outer saddle type sonic point recedes farther and the inner one proceeds 
inward. For a shock in accretion to be possible, pre-shock flow
parameter must lie somewhere on the branch $MD$ and the post-shock
flow parameter must lie somewhere on $MB$ as long as the energy and
entropy conditions: $E(x_{in}) \leq E(x_{out})$ and $s(x_{in}) \geq
s(x_{out})$ are satisfied. Similarly, for a shock in winds, the pre-shock
and post-shock flows must lie on the branches $AM$ and $MC$ respectively.
Typical Rankine-Hugoniot transitions (energy preserving)
are shown by horizontal arrowed dashed lines; difference in entropies
at arrow heads determine the entropy generated at the shocks. In the cases
where viscosity (heating) or $g$ (cooling) is non-zero, the considerations are 
the same, except that the transitions are not necessarily horizontal
even for Rankine-Hugoniot shocks, because the energy of the
flow at the two sonic points could be quite different. A 
typical such shock transition in the accretion $a \rightarrow a$ is
shown on the $\alpha_P=0.8$ curve (see, C89). 
For a VTF, the intersection (like $M$)
still separates two basic types of flows. As the viscosity and $f$ is
increased, the outer sonic point may no longer remain saddle type
and only the inner sonic point may exist. Thus shock transition may
no longer be possible. Flow parameters (e.g., the inner sonic point)
originally on the branch of type $AM$, move over to the branch of type
$MB$ (i.e., below $M$) as $\alpha_P$ is increased from $0$ 
to $\alpha_{c1}$. Thus for $\alpha_P <\alpha_{c1} (x_{in})$, the
flow will pass though the inner sonic point and join to a Keplerian disk at a
large distance ($x_{Kep}$ is determined by eq. 4a). These flows will stay on 
the branch $AM$ and can participate in shock-free accretion only and
not in accretion shocks. They can also take part in outwardly moving winds. 
If $\alpha_P>\alpha_{c1} (x_{in})$, the flow with the same 
$x_{in}$, will belong to the branch $MB$ and shocks become possible.
However, it escapes the shock region for $\alpha_P>\alpha_{c2} (x_{in})$.
Thus, for $\alpha_P >\alpha_{c2}$, the flow may pass through 
the inner sonic point without the shock, although, due to higher
viscosity $x_{Kep}$ is smaller (eq. 3a). 
It is to be noted that the dichotomy in topology 
in terms of the variation of $\alpha$ as discussed
here is valid only when $x_{in}$ and $l_{in}$ are held fixed. 
When $\alpha$ is held fixed, however, critical angular momentum
or critical sonic point location would be obtained which
would similarly separate the topologies.

The origin of the critical
viscosities is illustrated in Fig. 1b, where we plot two sets of 
curves, one for $\alpha=0$ and the other for $\alpha=0.01$ (marked) around the 
crossing point $M$. The curves marked ``inner''
and ``outer'' represent the quantities as the inner and the outer sonic points
are respectively varied. Heating efficiency factor $f=0$ is assumed for 
illustration. A flow which can pass through an inner sonic point marked ``A''
even without viscosity (C89) will approach the point 
``M'' (namely, in a zone which can produce shocks in winds) 
as viscosity is increased. For $\alpha_P=\alpha_{c1}$,
the point ``A'' coincides with ``M''. $\alpha_{c1}$ 
clearly has to depend on the location of ``A'' itself, namely,
the inner sonic point $x_{in}$ through which the flow must pass.
The inner sonic point $x_{in}$, in turn, depends on the specific
energy and cooling processes in the flow.
With further increase of viscosity, the flow having the same sonic 
point $x_{in}$ will slide down this region (and reach at ``a'', for example,
for $\alpha_P=0.01$) while passing through the zone of accretion-shocks 
(just below $M$). The point marked ``B'', which
was originally within the zone of accretion-shocks, would escape
to ``b'' for $\alpha=0.01$. This escaping process is the origin 
of $\alpha_{c2}$. It is easy to show that increasing 
the cooling parameter $g$ (or, decreasing heating
parameter $f$) has exactly the opposite effect on the quantities  
belonging to the inner sonic point. However, the 
branch representing the outer sonic point acts differently.
The flow originally passing through the  outer sonic point at ``C''
slides away (to ``c'') from the midpoint ``M'' as viscosity is increased. This 
point thus goes out of the region in which wind shocks could form (C89).
Similarly, the flow with outer sonic point at ``D'', which is capable of
participating in an accretion shock (C89) approaches the midpoint
``m" and soon cross over ``m'' so that accretion shocks can no longer
form for the flow passing through that particular outer sonic point. The effect
of increasing the cooling parameter $g$ (or, reducing the
heating parameter $f$) is also the same. Increasing $f$ brings back the
point ``$d$" into accretion shock region, and thereby increasing the
critical viscosity parameter for which shocks can form. These important
conclusions will be illustrated in the next section. 

It is to be noted that the actual values of $\alpha_{c1}$ and $\alpha_{c2}$
themselves are not only functions of the heating and cooling parameters
and the location of the sonic points (or, equivalently, energy density
of the flow), they also depend on the viscosity prescriptions that is
employed. Thus, e.g., critical viscosity parameters for $\alpha_P$
prescription would be roughly twice as much as for $\alpha_\Pi$
prescription. Similarly, it would be a little different if $W_{x\phi} (1)$
stress were used throughout. The only relevant point is that these two
crtical values exist which distinguishes flow topologies.

\subsection{$\Pi$-Stress}

In this case, the analysis is carried out 
with $W_{x\phi}=-\alpha_{\Pi} \Pi$. This is used
in computing the angular momentum distribution (3b). 
$Q^+$ is computed using MISStress prescription of Section 3.1.
$$
\frac{d\vartheta}{dx}= -\frac{N}{D} ,
\eqno{(9)}
$$
where the numerator $N$ is,
$$
N=\left [\frac{2n+1}{a}-\frac{4na\alpha_\Pi \omega}{(2n+3)\vartheta}
\right ] \left [ a\frac{5x-3}{2x(x-1)}+
\frac{\gamma}{a}(\frac{l^2}{x^3}-\frac{1}{2(x-1)^2}) \right ]
$$
$$
\ \ \ \ \ \ \ \ \ \ \ \ \ \ \ \ \ + \frac{5x-3}{2x(x-1)}
-\omega[\alpha_\Pi \vartheta (\frac{2n}{2n+3}\frac{a^2}{\vartheta^2}+1)
-\frac{2l^2}{x^2}]
\eqno{(10a)}
$$
and the denominator $D$ is,
$$
D=\left [ \frac{2n+1}{a}-\frac{4na\alpha_\Pi \omega}{(2n+3)\vartheta}
\right]\left [- \frac{\gamma}{a} (\vartheta-\frac{a^2}
{\gamma \vartheta})\right ] +\frac{1}{\vartheta}
-\alpha_\Pi \omega (1-\frac{2n}{2n+3}\frac{a^2}{\vartheta^2}).
\eqno{(10b)}
$$
Here,
$$
\omega=\frac{\alpha_\Pi f}{\vartheta}[\frac{2n}{2n+3}+\frac{\vartheta^2}{a^2}].
$$
The exact expressions for the Mach number and the sound velocity
at the sonic point $x_c$ are given by,
$$
M_c^2(x_c) = \frac{n+1-\alpha_\Pi^2 f \frac{2n(4n+5)}{(2n+3)^2}
+ [(n+1)^2-\frac{8n\alpha_\Pi^2 f (n+1)^2}{(2n+3)^2}
+\gamma^2 \alpha_\Pi^4 f^2 (
\frac{2n}{2n+3})^4 ]^{1/2}}{\gamma(2n+1)-\gamma\alpha_\Pi^2f(\frac{4n}{2n+3})
+\alpha_\Pi^2 f}
$$
$$
\ \ \ \ \ \ \ \ \ \ \ \ \ \ \ \ \ \ \ \ \ \approx \frac{2n}{2n+1} 
{\rm for \ \alpha_\Pi << 1 }
\eqno{(11a)}
$$
and
$$
a_c(x_c)=\frac{-{\cal B} + ({\cal B}^2 - 4 {\cal A}{\cal C})^{1/2}}{2{\cal A}}
\eqno{(11b)}
$$
where,
$$
{\cal A}={\cal D}
[\frac{5x_c-3}{2x_c(x_c-1)}+\frac{\gamma \alpha_\Pi^2}{M_c^2 x_c}+\frac{5x_c-3}{2x_c(x_c-1)}
+\alpha_\Pi^2 f (\frac{2n}{2n+3}+M_c^2)^2 ,
$$
$$
{\cal B}=2{\cal D}l_{in}\gamma\alpha_\Pi \frac{\frac{2n}{2n+3}+M_c^2}{x_c^2 M_c}
+2\alpha_\Pi f l_{in} \frac{\frac{2n}{2n+3} + M_c^2}{x_c^2 M_c} ,
$$
$$
{\cal C}={\cal D} [\frac{\gamma l_{in}^2}{x_c^3} -\frac{\gamma}{2 (x_c-1)^2}],
$$
and
$$
{\cal D}=(2n+1)-\frac{\alpha_\Pi^2 f}{M_c^2} (\frac{2n}{2n+3}+M_c^2)
\frac{4n}{2n+3}.
$$
These results also go over to the inviscid solutions
(C89) and isothermal solutions (C90a) in appropriate limits.
The nature of the sonic points and the behavior are very similar to
what is shown in Figs. 1(a-b). Differences occur in 
the angular momentum distribution of global solutions since 
$\Pi$-stress preserves angular momentum through shock waves as well.

Note that in both of these cases, the heating parameter $f$ always appears
along with the $\alpha$ parameter. This is because of our assumption
that the advected flux ($Q^+-Q^-$) could be written as $fQ^+$. This immediately
implies that, as if, cooling also disappears if $\alpha=0$. In general this
is not true: cooling can proceed independently of the heating process
depending on cooling rates which are functions the optical depth and
accretion rates. The transonic inviscid disks with bremsstrahlung
cooling alone which has this property is studied in MSC96. In this
work global solutions which passes through sonic points and shocks
are presented as functions of accretion rates.

Recently, Narayan \& Yi (1994)
considered a similar set of equations (1a)-(1e) and find global solutions 
using self-similar procedure in Newtonian potential. The solutions with 
$f\rightarrow 1$ were termed as `advection dominated'.
Since this treatment is self-similar, flow 
does not have any preferred length scale, such as sonic points,
or shock waves and therefore the conclusions derived from this
work are likely to be inapplicable or incorrect to describe
astrophysics around black hole or neutron stars (Narayan \& Yi, 1995ab;
Narayan, Yi \& Mahadevan, 1995; Narayan, McClintock, J.E. \& Yi, 1996;  
Lasota et al., 1996; Fabian \& Rees, 1995). 
For instance, our Fig.1(a-b) would not
simply exist in this self-similar treatment. The VTF
close to the horizon ($r \sim 10-100x_g$) must fall much
faster than a self-similar flow because the gravitational pull 
is stronger than a Newtonian star, and therefore emission properties
of the hard  component would be seriously affected, though soft
components which are emitted from regions far away from the black hole
should be less affected. Secondly, as we shall
show, the so-called `advection dominated solutions' do not constitute
any new class of solutions (indeed the terminology itself is unfortunate
since, as well shall see, the flow is actually always rotation
dominated close to the black hole, except, perhaps, just outside the
horizon. See, Fig. 7a below). We shall comment on other differences later. 

Before we present the global solutions, we wish to make a 
few comments about the viscosity prescriptions and the usefulness of one
over the other. In C90a and C90b, we have used both the cases where,
$W_{x\phi} (1) = -\alpha P$ and also where $W_{x\phi} (2) = \eta x d\Omega/dx$
and we found (also see, CM95) the results to be similar ($\eta=\alpha P h(x)
/\Omega_{Kep}$ is the coefficient of viscosity). In the above discussions 
we chose the first prescription
since it makes the angular momentum distribution completely algebraical
(eqs. 4a, 4b). An important corollary is that, we could now start
the integration by supplying the  integration constant $l_{in}$ and the
inner sonic point $x_c$ only and {\it derive} the location 
$x_{Kep}$ from which the low deviates from Keplerian disk. 

On the other hand, if the second prescription of the viscous stress
is used, the angular momentum distribution would be (C90a, C90b; CM95),
$$
l-l_{in}= \frac{\alpha_P P x^3 h}{{\dot M} \Omega_{Kep}}\frac{d\Omega}
{dx} .
\eqno{(12)}
$$
By virtue of the identification of $l_{in}$ to be the flow angular momentum
on the horizon, the flow automatically becomes shear free on the 
horizon. With this stress, it is easy to show that,
$$
\frac{d\vartheta}{dx}=\frac{N}{D} ,
\eqno{(13)}
$$
where, the numerator $N$ is,
$$
N=\frac{2n}{2n+1}\frac{5x-3}{2x(x-1)}+\frac{l^2}{a^2x^3}-\frac{1}
{2 a^2(x-1)^2} + \frac{f \gamma (l-l_{in})^2 \vartheta}
{\sqrt{2} \alpha_P a^4 x^{5/2}(x-1)(2n+1)}
\eqno{(14a)}
$$
and the denominator $D$ is,
$$
D=\frac{1}{\vartheta}(\frac{\vartheta^2}{a^2}-\frac{2n}{2n+1})
\eqno{(14b)}
$$
The first term is related to the geometric compression of the
flow, second and the third terms represent a competition between
the gravity and centrifugal force. The fourth term (containing $f$ is the
contribution from the heating/cooling effects.
This term comes separately in the numerator exactly as in the case of
bremsstrahlung (MSC96) except that in latter case the term appears
with an opposite sign consistent with cooling in presence of
weak viscosity ($f<0$). Otherwise the condition $N=0$ here
is not very helpful since it requires a knowledge of
$l(x_c)$ which is itself a priori not known (eq. 12). The expression
is consistent with $\alpha\rightarrow 0,\ l \rightarrow l_{in}$
or, $x\rightarrow 1, \ l \rightarrow l_{in}$ and thus consistent with
inviscid solution of C89 when $f=0$ is chosen. 
The denominator gives the Mach number
at the sonic point exactly as obtained for inviscid case (C89).
One can obtain similar expressions when
$\eta$ in $W_{x\phi}$ is written in terms of the total pressure
(ram plus thermal, eq. 7b of CM95). 

This prescription, however, poses a few difficulties: (1) one has to solve
one extra differential equation (eq. 12) for angular momentum; (2) one
no longer has an algebraic condition at the sonic point  
and therefore study of critical point behavior is difficult
(Flammang, 1982), and finally,
(3) one definitely has to start the integration from the outer 
edge $x_{Kep}$ of the flow, particularly when one is using
pseudo-Newtonian potential. The problem (3) is severe since it would 
not be known a priori whether the flow would go through the sonic point
for a given choice of outer boundary condition, or even if it 
is forced to go, whether the derivatives at the sonic point
would be continuous. Because of these reasons, we have chosen 
the first prescription ($W_{x\phi} = -\alpha_P W$
or $W_{x\phi}=-\alpha_\Pi \Pi$) to consider the angular momentum distribution
while adopting MISStress prescription for the cooling term.

If the cooling term were chosen using $W_{x\phi}=-\alpha_P  P$ prescription, one
would have $Q^+$ as:
$$
Q^+=\frac{W_{x\phi}^2}{\eta} = 
\alpha_P P h(x) \Omega_{Kep} \sim \alpha_P P a / \sqrt{\gamma}
\eqno{(15)}
$$
the sonic point analysis becomes more simplified. 
The general result, however, remains qualitatively the same.

No matter what prescriptions are used, the positivity of the sound
speed at the sonic point (obtained from the vanishing condition $N=0$)
requires that the angular momentum at the sonic point be sub-Keplerian
(cf. eq. 8b, 14a).
This was pointed out by Abramowicz \& Zurek (1981) in the
context of adiabatic accretion (see, C90b). 
We prove in this paper that any transonic disk 
is necessarily sub-Keplerian at least in some region at and
near the sonic point,
provided the advective term $Q^+-Q^- > 0, i.e., f\gsim 0$. Of course,
when the flow is super-cooled ($f<0$) it {\it can} be sonic even 
in a super-Keplerian flow depending on the competition between the
geometric heating  factor and  the cooling factor.

\section{SOLUTION TOPOLOGIES}

To obtain a complete solution, one must supply the boundary
values of energy (or, accretion rate) and angular momentum
for a given type of viscosity ($\alpha$) and cooling parameter ($g$).
This is analogous to Bondi solution where only one parameter, namely,
energy density or accretion rate is required. Instead of supplying above 
mentioned quantities, we supply here one sonic point location $x_c$ and the 
angular momentum constant $l_{in}$ (Eq. 3). The equations
are integrated from the sonic point inward as in C90a and C90b till
they reach the black hole horizon or the neutron star surface. Similarly,
they are integrated outward till the Keplerian distribution is achieved.
In the case of neutron star accretion, one could supply the outer
sonic point instead if the star is big enough to engulf the
inner sonic point within its surface.
Without any loss of generality, we choose the cooling parameter
$g$ (i.e., $f$ for a given $\alpha$) to be a constant in the analysis below.

Our choice of initial parameters $x_c$ and $l_{in}$ stems from the 
following considerations: since a black hole accretion is transonic
(C90ab), the flow  has to pass through `A' sonic point at $x_c$.
Secondly, since we want the flow to originate presumably from a Keplerian
disk, it has to carry some angular momentum $l(x_{Kep})$, 
a part of which would be transported away by viscosity and the other
part must enter through the horizon.
Exact amount of entry of angular momentum $l_{in}$
is not of much concern (unless one is interested in the spin-up process
of a black hole). Thus, $l_{in}$ and $l_(x_{Kep})$ are
related through eq. (3) and we could have, in principle, supplied
$x_{Kep}$ instead. But this is very much uncertain (and is physically
unintuitive) as it could vary anywhere from $10$ 
to $10^6x_g$. On the contrary, the acceptable range of 
the angular momentum of the accreting solution at the inner edge 
is very small (from, say, $1.5$ to $\sim 2$, see, C89, C90ab). Thus, our 
approach has always been to choose the angular momentum at the
inner edge as a free parameter, and then integrate backward to see
where the flow deviated from a Keplerian disk (i.e., what angular
momentum the flow started with) in order to have  
$l_{in}$ on the horizon. In other words, in our approach, $x_{Kep}$
is the eigen value of the problem. 
As in C90a and C90b, in what follows, we shall use this approach as well.

\subsection{General Behavior of Globally Complete Solutions}

Figs. 2(a-b) show examples of global solutions passing through
the inner sonic points. In each small box in Fig. 2a, we plot Mach number 
$M=\vartheta/a$ (vertical scale goes from $0$ to $2$) as function of the
logarithmic radial distance (scale goes from $0$ to $50$). On the upper axis,
we write $\alpha_\Pi$ parameters (marked as $0.0001$, $0.05$, $0.2$ and 
$0.4$) and on the left axis we write the heating 
parameter $f$ (marked as $0.0$, $0.1$, $0.5$, and $1.0$).
Each of the grid number of the  $4\times 4$ matrix that is formed is written
in the upper-left corner of each box. In Fig. 2b, we show the ratio $R$ 
(vertical scale goes from $0$ to $1.5$) of disk angular momentum to the 
Keplerian angular momentum to emphasize on the degree at which
the flow is non-Keplerian. The short-dashed horizontal lines in each box is 
drawn at $R=1$. Other parameters fixed for the figures are $x_c=2.795$,
$l_{in}=1.65$ and $\gamma=4/3$. In these figures, the sonic points
are not located at $M=1$ but at an appropriate number computed
assuming corresponding polytropic index,
the viscosity and heating parameter as given in eq. (11a).

Whereas in all these figures the sonic point is saddle type, whether
or not the flow will participate in a shock or remain shock-free will depend
on its global topology. In the first column of Fig. 2a, the flow is almost
inviscid and the results are almost independent of the heating parameter $f$.
The flow joins with the Keplerian disk at several thousand Schwarschild radii 
(outside the range of Fig. 2b, but see, Fig. 7b below). The flow leaves the 
Keplerian disk and enters the black hole straight away through the 
inner sonic point. This open topology is the characteristics of the parameters
chosen from the branch $AM$ of Fig. 1a, i.e., the inflow can pass through
inner sonic point without a shock, or, an outflow will form (with or without
a shock), depending on where on the branch $AM$ the parameter is located.
In the second column, the viscosity is higher, and the topologies are
closed. This implies that $\alpha_\Pi >\alpha_{c1}$ is already reached
and the same inner sonic point brought the flow from the branch $AM$
to $MB$. The angular momentum of the flow cannot join a Keplerian
disk unless a shock is formed, or the flow is shock free, but passes
only though the outer sonic point if it exists. For lower heating parameter
the flow topology opens up again when viscosity parameter
is further increased (column 3) and the flow again joins with the Keplerian
disk, but only at a tens of Schwarzschild radii (Fig. 2b). In this case,
$\alpha_\Pi >\alpha_{c2}$ is achieved and the flow topology
leaves the accretion shock regime on the branch $AM$. For a higher heating
parameter $f$, $\alpha_{c2}$ is higher, if the sonic point still remains
of saddle type. This is consistent with our understanding (Fig. 1b) that 
increasing $f$ (or, reducing $g$) brings back the flow into the shock regime
for a given viscosity parameter.  Note that as $f$ crosses, say, $0.5$,
i.e., as the cooling become more inefficient, the integral curves change their
character: the spiral with an open end goes from clock-wise to anti-clockwise.
The implication is profound. For $f\lsim 0.5$ the closed spiral 
surrounding the 
open spiral can still open up to join a Keplerian disk (e.g., grids 13, 23),
but for $f\gsim 0.5$ closed spiral can no longer join with a disk (e.g., grids 34,43,
they could, in principle open up to a `Keplerian wind'!, see also C90a,b).
Thus we prove that only for higher cooling ($f\lsim0.5$) and higher viscosity 
($\alpha>\alpha_{c2}$), Keplerian disk can extend much closer to the black 
hole, otherwise it must stay much farther away and the transonic advective solution 
will prevail.

The change of topologies by a change of viscosity is not surprising.
Increase in viscosity increases angular momentum at the sonic points.
At smaller viscosities, the sub-Keplerian flow becomes Keplerian
very far away and as we discussed before (above eq. 3), all the
three sonic points could be present. At a high enough viscosity,
flow becomes Keplerian very quickly and only one (the inner)
sonic point is possible.
(Note that we mean the distance from the {\it horizon}
when we use the phrase `far away' or `quickly'.)

Though we shall discuss in detail in section \S 6, we like to
point out the important result that the flow with a lower viscosity
and higher cooling joins with the Keplerian disk at a farther
distance than the flow with a higher viscosity. 
This implies that
a disk with a differential viscosity with lower viscosity
at higher elevation can simultaneously have a Keplerian disk on the
equatorial plane and a sub-Keplerian disk away from the equator.
This has already been observed in isothermal disks (C90b, CM95) and 
this consideration has allowed us to construct the accretion disk
of more general type (C94, C96a). This also allowed us to
obtain the most satisfactory explanation, to date, of the observed
transition of soft and high states of galactic black hole candidates 
(CT95, Ebisawa et al, 1996). A similar picture of accretion flow is obtained
when  $\alpha$ is kept fixed (even increasing vertically upward)
for the entire disk, but $l_{\rm in}$ or $x_{\rm in}$ also increases 
away from the equatorial plane. One requires very special
cooling efficiencies to fulfill these constraints.

We continue to emphasize the importance of the understanding
of the nature of the inner sonic points, by varying its
locations as in Fig. 3a. We mark $x_{in}$ in each box and keep
other parameters fixed: $\gamma=4/3$, $\alpha_\Pi=0.05$, and $f=0.5$.
As the sonic point location is increased, the open topology of the flow
(in branch $AM$) becomes closed (in branch $MB$) and ultimately the
physical solution ceases to exist as the inner sonic point no longer 
remained saddle type (cf. Fig. 1a). The only available solutions remain
those passing through the outer sonic point (discussed later). Note that the
inner sonic point continues to remain saddle type even when it
crosses $r=3x_g$, i.e., marginally stable orbit. This is because
the marginally stable orbit in fluid dynamics (i.e., in presence of
pressure gradient forces) does not play as much special role
as in a particle dynamics. It is easy to show that the similar crossing
at $r=3x_g$ takes place for a large range of $\alpha$ parameters.
In Fig. 3b, we vary angular momentum $l_{in}$ (marked on the upper axis)
and the heating parameter $f$ (marked on the left axis) while
keeping $x_{in}=2.8$ and $\alpha_\Pi=0.05$. We note that for
very low angular momentum, the flow behaves like a Bondi flow,
with only a single sonic point. As angular momentum is
increased the topology becomes closed and the flow can enter the
black hole only through a shock or through the outer sonic 
point if it exists.

So far, we discussed the nature of the inner sonic point. However,
as in an inviscid polytropic flow (Liang \& Thomson, 1980; C89), or,
isothermal VTF (C90a, and C90b), the general case that is 
discussed here also has the three sonic points as is obvious in
Fig. 1a. In Fig. 4, we show Mach number {\it vs.} logarithmic distance
when $x_{out}=35$ is chosen (with the same scale and other 
parameters as in Fig. 2a). In the first column, with low viscosity,
the flow either passes through the outer sonic point only,
or can pass through a shock and subsequently through the inner sonic point
(if the shock conditions are satisfied). 
As the viscosity is increased, the topology is closed and the flow
parameters must be different so as to allow the flow through 
an outer sonic point ($x_{out} > 35$) which has an
open topology and which smoothly joins with the Keplerian disk farther away.
It could also subsequently pass through the inner sonic point if shock 
conditions are satisfied. The choice must finally depend upon the 
the flow parameters as illustrated in Fig. 1a. Note that, in general,
it is difficult to have a saddle type outer sonic points for higher
$\alpha_\Pi$, partly because it is defined to be about half of $\alpha_P$
and partly because the outer sonic point itself recedes at viscosity
is increased (Fig. 1a) and therefore flow does not pass through
a given outer sonic point if the viscosity is raised.

\subsection{Solutions which Contain Shock Waves}

In Fig. 5a, we present Mach number variation with the logarithmic
radial distance. The Rankine-Hugoniot conditions (namely,
conservation of the mass flux, momentum flux and energy flux
at the shock front) in the vertical
averaged flow (C89) were used to obtain the shock locations.
The flow parameters chosen are $x_{out}=50$, $l_{in}=1.6$,
$\alpha_\Pi=0.05$, $\gamma=4/3$ and $f=0.5$. The shock
conditions in turn force the flow to have a shock at $x_{s3}=13.9$
(using notation of C89) and to pass through the inner sonic
point at $x_{in}=2.8695$. This location is computed by equating
its entropy with the amount of entropy generated
at the shock and subsequently advected by the flow plus the
entropy generated in the post-shock subsonic flow
(similar to C90a,b where energy advection conditions were considered).
The shock itself (shown here as the vertical transition with a single
arrow) is assumed to be thin and non-dissipative, i.e., energy
conserving. In presence of viscosity, the shock would be
expected to smear out. Thus $x_{s3}$ calculated assuming infinitesimal
shock width represents an `average' distance of the shock
from the black hole. We have shown also a double-arrowed vertical transition
where a shock will form in a neutron star accretion with subsonic
inner boundary condition and the flow locking in with the surface.
It is easy to verify that the shock conditions are
satisfied at $x=2.39$. A neutron star of mass $1.4M_\odot$ and radius 
$r_*=10$ km (i.e, $r_*=2.38 r_g$) will marginally fit within the 
shock. In case the star surface is bigger than the
inner sonic point, the post(single-arrowed)shock branch will be completely
subsonic, and not transonic as is shown here. We have also drawn only 
$x_{s3}$. The other location $x_{s2}\sim 4$, closer to the inner sonic
point is unstable as will be shown below.

In order to show that the shock transitions shown in the above Figure
are real, and stable, we show in Fig. 5b, not only the Mach numbers
of the subsonic and supersonic branches, but also other physical
quantities along these branches. The solid curves represent the branch 
passing through the outer sonic point located at $x_{out}=x_c=50$ and 
the long dashed curves represent the branch passing through the
inner sonic point at $x_{in}=x_c=2.8695$. 
The flow chooses this subsonic branch for $x<x_{s3}$ 
since the entropy of the flow is higher at the inner sonic point (C89).
We have plotted the shock invariant (C89) function appropriate for
a vertically averaged flow,
$$
C= \frac{\left[ M(3\gamma -1) + \frac{2}{M}\right ]^2}
{2 + (\gamma -1 ) M^2 },
\eqno{(16)}
$$
the angular momentum distribution $l(x)$ ($\times 2$),
the Mach number distribution $M(x)$, the total pressure $\Pi$
(in arbitrary units), the local specific energy $E(x)$, radial velocity and
the proton temperature $T=\mu m_p a^2 /\gamma k$ (in units of $2 \times
10^{11}$K). Here, $\mu=0.5$ for pure hydrogen, $m_p$ and $k$ are proton mass 
and Boltzmann constant respectively. At the shock, the temperature goes up and
the velocity goes down, in the same way as in our earlier studies.
The solid and dashed curves describing $C$, $\Pi$, $E(x)$
variations intersect at the shock $x=x_{s3}=13.9$, consistent with the 
Rankine-Hugoniot condition. Though the total pressure $\Pi$ in these two 
branches intersect at two locations suggesting two shocks (as in C89, C90ab)
only the one we marked is stable, as can be easily verified by
a perturbation of the shock location (CM93).
At $x=x_{s3}=13.9$, if the shock is perturbed outwards, the pressure
along the pre-shock flow (solid curve) is higher than that along
the post-shock flow (dashed curve) (Fig. 5b). Thus the shock is pushed
inwards. Similarly, if the shock is perturbed inwards, it is
pushed outwards due to higher pressure in post-shock flow.
This is not true for the intersection at $x_{s2} \sim 4$ and therefore
the shock solution at $x_{s2}$ is unstable.
The location of the shock on the neutron star accretion is
also obtained in the same way, but this time one has to compare
quantities of the super-sonic branch passing through the 
outer sonic point, with the quantities of the sub-sonic branch
at $x<x_{in}$ (Fig. 5a).

We discussed only about those shock transitions which do not instantaneously
release energy or entropy at the shock locations. If they do, the
shock conditions have to be changed accordingly (Abramowicz \& Chakrabarti,
1990; C90b) and the shock locations appropriately computed.

So far, we chose only $f=$constant solutions.
One can always choose a suitable function $f(\alpha, x, {\dot M})$ which
satisfies $f \rightarrow 0$ for $x \gsim x_{Kep}$ and $f\rightarrow 0.5-1$
(depending on cooling efficiency) for $x\sim x_c$
and redo our exercise. This will clearly be the combination of 
results presented above where outer sonic point is chosen for $f=0$ and 
inner sonic point is chosen for $f \sim 0.5-1$. A cooling function $g(\tau)$
to accomplish this is already presented in Chakrabarti \& 
Titarchuk (1995). Work with actual 
heating and cooling is in progress, and we shall report them in future.

The difference in the boundary conditions in black hole and neutron star
cases give rise to an important observational effect. The bulk motion
of the optically thick converging inflow (Blandford \& Payne, 1981)
could `Comptonize' soft photons through Doppler effect to produce a hard 
spectra of slope $\sim 1.5$ (Chakrabarti \& Titarchuk, 1995) observed in the
galactic black hole candidates (e.g., Sunyaev et al., 1994). In a neutron
star accretion the flow is subsonic close to the
surface and such a power law is neither expected nor observed.

In Fig. 6, we present a montage of solutions involving the shock waves
for $\gamma=4/3$. ($\alpha_\Pi, \ f$) parameter pair is written in each box.
Mach number is from $0$ to $2$ and the radial distance
is varied from $0$ to $100$. The outer sonic
points are located at $x_{out}=50$, and the inner sonic points
were determined from the evolution of the flow after the shock, 
are also shown. In the boxes containing only the flow from outer
sonic point, shock conditions were not found to be satisfied
in a black hole accretion. We therefore did not draw the
branch with inner sonic point, since it would be meaningless to
do so. The shocks in black hole and neutron stars for
($\alpha_\Pi, \ f$) parameters ($0.07,\ 0.1$), ($0.05,\ 0.3$),
and ($0.05,\ 0.5$) are located at $15.025$ and $2.38$,
$10.35$ and $2.34$, and $13.9$ and $2.392$ respectively.
It is to be noted that the self-similar solutions in Newtonian
potential (e.g., Narayan \& Yi, 1994) shocks cannot form since
no length scale is respected by self-similarity assumption and
the flow always has constant Mach number and does not pass through
sonic points of any kind.  It is to be noted that for accretion around
a neutron star, two shocks (of type $x_{s1}$ and $x_{s3}$
in C89 notation) may form if the star is compact enough. This would
make computation of a neutron star spectra more complicated.

\subsection{Advection vs. Rotation, Keplerian vs. Non-Keplerian}

As in the past, we define the flow to be advection dominated
when $\vartheta > v_\phi=l/x$ and rotation dominated when
$\vartheta < v_\phi=l/x$. It is interesting to
study whether the flow is dominated by the advection or rotation,
as the flow starts deviating from the Keplerian disk.
In Fig. 7a, we present the ratio $\vartheta/v_\phi$ of a few solutions
already presented. The solutions with labels `11', `12' and `13' 
are from the first row and the solution with label `43' is from 
from the third column of Fig. 2a. The solution labeled `shock' corresponds
to the case presented in Fig. 5a. Except those marked
`12' and `43', other solutions smoothly match with the Keplerian disk
at the outer edge
as they become more and more rotation dominated. Advection domination
starts much closer to the black hole, although, interestingly, flow again 
becomes rotation dominated as it comes closer to the black hole. At the shock,
the flow goes from advection dominated to rotation dominated although
the angular velocity itself is continuous (Fig. 7b). The solutions
marked `12' and `43' are either to be joined by shock waves 
(i.e. a flow first passing through the outer sonic point) or, are not possible
at all, since they do not by themselves smoothly join with a Keplerian disk.

In Fig. 7b, we present the ratio of disk angular momentum to the
local Keplerian angular momentum as a function of the logarithmic
radial distance. The curves are labeled by viscosity parameters
$\alpha_\Pi$. Other parameters of the group labeled `A' are:
$l_{in}=1.88$, $x_{in}=2.2$, $\gamma=4/3$, $f=0$ and of the group
labeled `B' are $l_{in}=1.6$, $x_{in}=2.8695$, $\gamma=4/3$, $f=0.5$. 
Note that for a given set of flow parameters, as the viscosity is reduced,
the location $x_{Kep}$ where the flow becomes Keplerian is also increased
(eq. 4a; C90a,b). Secondly, the flow can become super-Keplerian close
to a black hole, a feature 
assumed originally in modeling thick accretion disks (e.g. Paczy\'nski
\& Wiita, 1980). Thirdly, if the viscosity is high, the flow may become
Keplerian immediately close to the black hole. This behavior
may be responsible for hard state to soft state transitions in black hole
candidates as well as novae outbursts which are known to depend
on viscosity on the flow. We discuss this in the final section. 
Note that the angular momentum distribution of the curve marked $0.05$ is 
continuous even though it has a shock wave at $x_{s3}=13.9$ (Fig. 5a-b).

Since the thick accretion disks are traditionally considered to be
those which are sub-Keplerian and at the same time rotationally dominated,
we find  from Figure 7b that there are essentially {\it two} thick 
accretion disks, one {\it inside} the other, so to speak. One
is a `big' thick disk, whose outer edge starts where the flow
deviates from Keplerian, and the other is a `small' thick disk,
which occupies the post-shock flow. Our thick disks are more
accurate than traditional thick disks, because we include advection as well.

\subsection{Dependence on Polytropic Index}

So far, we have discussed the solution topologies, with the polytropic 
index $\gamma=4/3$. In general, the index could be higher or lower, depending on the
radiation and magnetic field content, and for a fully self-consistent solution
one is required to compute this index as the flow evolves, rather
than choosing it a constant. This is beyond the scope of the
present analysis. In C90a and C90b, we considered $\gamma=1$ results which
would be used for both optically thick or thin advective flows.
We now give some flavor of
solution topologies when the extreme case of $\gamma=5/3$ is chosen.

Fig. 8 shows the energy-entropy plot for $\gamma=5/3$ and $l_{in}=1.65$.
This is to be compared with Fig. 1a, where $\gamma=4/3$ was used instead.
Other notations are identical.  The arrow indicates the 
variation as the sonic point is increased.
The important point to note is that, in this case, the outer saddle type
sonic point does not exist. In other words, the flow must
adjust itself to pass through the inner sonic point alone (shown
by the solid curves). A corollary of this is that the flow will not have
a centrifugally supported shock wave, as discussed here, unless the
flow is already supersonic at the outer boundary (or, the flow geometry
is different, e.g., MSC96) or the shock is not centrifugally supported
but forms due to some other effects (such as, external heating; see, e.g.
Chang \& Ostriker, 1985). It is easy to show that similar
absence of the outer sonic point prevails 
for $\gamma \geq 1.5$ (see, Fig. 3.1 of C90b).
It is possible that relativistic flows close to a black hole has 
$\gamma\sim 13/9$ (e.g., Shapiro \& Teukolsky, 1983), thus may be the
possibility of shocks are more generic.

Finally, for the sake of completeness, we close this section by showing 
the behavior of the solution topologies for $\gamma=5/3$. In Fig. 9a,
we vary viscosity and heating parameters as in Fig. 2a. Other parameters
are: $x_{in}=2.8$, $l_{in}=1.65$. With these choices, the topology is
closed even for an inviscid flow. As the viscosity is increased,
the topology opens up and joins with a 
Keplerian flow. If we started with an initially open topology,
similar to $\gamma=4/3$ case, we would have two critical viscosities
causing similar topological changes as in Fig. 1a. In Fig. 9b,
we show the topologies when heating parameter $f$ and
the angular momentum $l_{in}$ (marked on the left axis)
and the location of the inner sonic point $x_{in}$ 
(marked on the upper axis) are varied.

\section{ON THE COMPLETENESS OF THE GLOBAL SOLUTIONS}

Using simple combinatorics, we briefly argue here that there could be 
no other topologically distinct VTF solution other than what 
we described in this paper. For a physical solution, the final sonic point
through which the flow must pass (either outer or inner) just before entering
a black hole should be of saddle type.
Let us denote them by $S_+$ (positive slope) and $S_-$ (negative slope)
solutions respectively. In the inviscid case, central sonic point is center
type or `O' type (say, $O$ for convenience) which splits into two 
spiral type solutions which may be clockwise or anti-clockwise
when viscosity, heating and cooling are added. Let us denote them by 
$P_+$ for clockwise spirals and $P_-$ for anti-clockwise spirals.
Clearly, the following combinations of these sonic points (from inner edge 
outwards) form an exhaustive set:$S_+S_-$, 
$S_+S_-OS_+S_-$, $S_+(S_-P_-)(P_-S_-)S_+$,
$(S_+P_+)(S_-P_-)S_+S_-$,$S_-(S_+P_+)(P_+S_+)S_-$, 
$S_+S_-(P_+S_+)(P_-S_-)$.
These handful of choices are dictated by the fact that a saddle type solution
with a positive slope can only join with a clockwise spiral (this `joining'
is indicated by parenthesis) and a saddle type solution with a negative
slope can only join with an anti-clockwise spiral. Except the second case
with `O' type point, which we found in the inviscid flow (C89), the rest
have been shown in various figures of the previous section and C90a and C90b.
Although we consider only vertically averaged
flow here, the topologies are not expected to change
when a `thick' quasi-spherical flow is considered.
Indeed, the presence of shocks would be more generic
as they would appear even outside the 
equatorial plane because of weaker gravity. Similarly, when a Kerr
black hole is used, the centrifugal barrier become stronger
with the increase of the Kerr parameter because the horizon gets smaller. 
Thus, the shocks are formed for much wider parameter range in Kerr geometry. 

A fundamental assumption which allowed us to simply
classify these solutions is that the radial forces involved in the
momentum equation (eq. 1a) 
are `simple enough' so that the velocity variation of the
flow could still be reduced into the form (6) or (9)
since both the numerator and the denominator are only {\it algebraic}
functions and do not involve differential operators.
In the present context, this was possible by choosing $W_{x\phi}=
-\alpha P$ prescription of shear stress. Even then, if the
force were more complicated, as in the case of cooler wind solutions from
mass lossing stars where radiative acceleration term with 
nonlinear dependence of velocity gradient is included
(Castor, Abbott \& Klein, 1975), it would not be possible to 
reduce the governing equations into a first order differential
equation (vital to the  discussions of Bondi-type solutions). 
Critical curves, instead of critical points would
be present (Flammang, 1982; C90b) solution topologies of which would be more
complex. Discussion on this is beyond the scope of the present analysis. 

\section{DISCUSSION AND CONCLUSIONS}
 
In this paper, we presented for the first time the global solutions
of transonic equations in presence of viscosity, advection, rotation,
{\it generalized} heating and cooling. Our VTF solution starts from a
Keplerian disk at the outer edge and enters through the
horizon after passing through sonic point(s).
As in our earlier studies of inviscid (C89)
and isothermal VTFs (C90a and C90b), we emphasized here the
possibility of the formation of the shock as well where
two transonic solutions are joined together by means of
Rankine-Hugoniot conditions. Though we have used
general considerations of heating and cooling, we note
that no new topologies of solutions emerge other than what are discussed 
in C90a and C90b. However, unlike in C90a and C90b, where critical viscosities
are studied only in the context of shock formation, our detailed study
here indicates that there are indeed two types of critical
viscosities both of which depend on the parameters of the flow.
For $\alpha<\alpha_{c1}$,
accretion through inner sonic point is only allowed (i.e., no
shock) or winds with or without shock is allowed. In this case,
the flow joins with a Keplerian disk very far away (Fig. 2a).
For $\alpha_{c1} <\alpha <\alpha_{c2}$ the flow can have shocks if shock
conditions are satisfied, else the flow will pass through the
outer sonic point. For $\alpha>\alpha_{c2}$, the
flow will pass through inner sonic point again. In this case,
the flow joins a Keplerian disk very close ($x\sim 10 x_g$)
to the horizon. Whether both of these
critical viscosity parameters exist will depend on the flow
parameters, such as the location of one sonic point, the angular
momentum at the inner edge $l_{in}$ and the heating parameter $f$.
Once we specify these quantities, the entire solution topology,
including the shock location (if present), and the location 
where the flow joins with a Keplerian disk are completely determined.
Our results depend on the accretion rate through the cooling parameter $g$
(or, equivalently, through $f$) and always produce stable branch of the
solution in the ${\dot M}-\Sigma$ plane. This is possibly because of our 
choice that the cooling could be written as a constant fraction of heating term.
We also show that when the polytropic index is
higher than $1.5$, in a vertically averaged flow model 
the outer sonic point does not exist (though it exists if the disk is
thinner, see, MSC96), and therefore, shocks are possible
only if the flow at the outer boundary is already supersonic.
Since the total pressure $\Pi$ (and not
the thermal pressure) is continuous across the shock waves,
we used the $\Pi$-stress prescription (CM95) to study shock waves.
This prescription is always valid, and we recommand
its uses for the study of astrophysical flows around
black holes and neutron stars, where the advection effect, and therefore
ram pressure is important.

Since our VTF solutions have one {\it less} free parameter, the
${\dot M} (\Sigma)$ relation is monotonic with positive slope and always
represents the stable solutions. In other words, the thermal/viscous
instability is removed completely by the addition of advection
effects. This is true whether or not the flow contains a shock
wave. When more general cooling law
is used (with power law dependence on accretion rates, for example),
it is possible that our stable solutions could be `destabilized' in some
parameter space spanning ${\dot M}, \Sigma$ as originally
discussed by Meyer \& Meyer-Hofmeister (1983) (see also, Cannizzo, 1993). 
Thus, the problem with our model is not `how to stabilize the inner edge
of a Keplerian disk' but rather, `how to destabilize the perfectly stable
transonic disks', if indeed, the novae outbursts are signatures of such
instability. Similarly, because of our choice of cooling law $Q^- \propto 
Q^+$, we are unable to show direct influence of accretion rate onto to the
cooling efficiency. Our parametrization indicates that for {\it any} 
accretion rate, higher or lower efficiency of cooling is possible, 
just by changing the viscosity parameter. 
Furthermore, we find that the variation of $x_{Kep}$ with
$\alpha$ could be achievable by varying  $x_{in}$ and $l_{in}$
as well (Fig. 3) even when the entire flow has constant viscosity
parameter $\alpha$.
These questions are easily answered by assuming exact cooling
laws as far as possible. We shall examine these solutions in near future.

Unlike the properties of the more complex VTF
models, where non-linear radiative accelerations play major
roles, we used simpler momentum equation, relevant for hot
flows close to a compact object. This enabled us not only to
obtain global solutions in general, but also to discuss about the complete
set of topologies. Indeed, the same topologies were seen in isothermal
VTFs (C90a and C90b), and we argued that no other type of solutions are
possible either. The same conclusion holds
even when more general cooling laws (such as with power law dependence
on density and temperatures) are employed (MSC96). Our results have the
same accuracy as that of the description of the pseudo-Newtonian potential
of a Schwarzschild geometry. But
as we discussed in the introduction, no new topologies have 
been discovered for inviscid flows when full general relativistic models 
are solved (C96b). However, under some circumstances
the torque can become negative and the angular momentum is transported
inward. This happens just outside the horizon and does not affect the
solution topologies.

Having not discovered any new topologies, we believe that the unified 
scheme of global solutions presented by us (C93, C94, C95, CM95) with the 
knowledge of isothermal disks remains valid even for disks with general 
heating and cooling. In C94, we wrote ``... These 
findings are very significant as they propose a unifying
view of the accretion disks. This incorporates two extreme
disk models into a single framework: for inviscid disks,
strong shocks are produced, and for disks with high enough
viscosity, the stable shock disappears altogether and angular
momentum can become Keplerian." Our present grand unified 
global solution describes the most general form of accretion which goes 
over to the other disk models presented in the literature. This is 
not surprising, since we exactly solve the most general equations. Our disk
can pass through either or both the sonic points (when shocks are present)
while joining smoothly with the Keplerian disk at a distance $x_{Kep}$.
The post-shock flow, where the disk is rotationally dominated (Fig.
7a) behaves exactly as the thick accretion disks (e.g. Paczy\'nski \&
Wiita, 1980; see MLC94). If the shock does not form, the sub-Keplerian
optically thin flow will behave similar to the ion-supported
tori (Rees et al. 1982) because cooling is inefficient in an
optically thin flow, and the entire energy remains conserved
and is advected away (C89). 
However, our solution is more consistent than a conventional thick
disk model as we include advection term as well, and obtain 
the global solutions (see also related discussions in CT95).
In MLC94, we have already verified that the
results of vertically averaged solution (C89) are sufficiently
accurate. Our solutions indicate that winds may be produced
from the inflow with positive energy (C89; Fig. 1a) which are
verified through extensive numerical simulations (MLC94, Ryu et al. 1995). 
A knowledge of the dependence of $x_{Kep}$ on viscosity enabled
us to construct the most general form of accretion disk,
in which Keplerian disk in the equatorial plane is flanked by sub-Keplerian
flows above and below. Post-shock hot matter Comptonizes soft-photons
from the Keplerian disk to produce hard X-rays (CT95). This understanding
has resolved the long standing problem of transition of states of black hole
spectra. Furthermore, MSC96 shows that shock oscillations could be
responsible for the quasi-periodic oscillation. The difference of
the inner boundary condition in the neutron star and black holes
has resulted in a difference in spectral index from the emergent
spectra in soft states (CT95). Our most general solution
can also provide explanations of more complex phenomena, such as the
spectral evolution observed during novae outbursts (Ebisawa et al.,
1996). For instance, during the
quiescence stage, viscosity being low, the disk deviates from being Keplerian
farther away (Figs. 2a, 7a, 7b) and the generally optical radiation
is accompanied by very weak X-rays which are produced due to reprocessing
of the soft radiation intercepted by the sub-Keplerian inner disk. As the
viscosity at the outer edge is increased due to piling up of matter
(e.g., Cannizzo, 1993), the sub-Keplerian flow first rushes in close to the
black hole (since the in fall time scale is shorter for a sub-Keplerian
flow) increasing the hard X-ray component as commonly observed in the 
pre-outburst phases (Ebisawa et al., 1994). Subsequent increase
in Keplerian matter close to the black hole increases soft-X-ray component
as is observed. Thus, the present {\it grand unification}
of the accretion solutions is more complete, and successfully
bridges the gap between the well studied spherical Bondi flow
(e.g., Shapiro, 1973a,b; Ostriker et al., 1976; Begelman,
1978; Colpi, Maraschi \& Treves, 1984; Wandel, Yahil \& Milgrom, 1984; 
Blondin, 1986; Begelman \& Chiueh, 1988; Park \& Ostriker, 1989; Park 1990; 
Nobili, Turolla \& Zampieri 1991) and the Keplerian 
disks of Shakura \& Sunyaev (1973) and Novikov \& Thorne (1973).
An important ramification of having sub-Keplerian disks is that the 
azimuthal velocities are less than that of a Keplerian flow. 
The masses of the
central object determined from such considerations (e.g., using Doppler
shifts of emitted lines) are naturally higher (Chakrabarti, 1995)
than what it would have been if Keplerian motions were assumed instead.
Similarly, due to inefficiency of emission processes in disks with
advection, the masses of the
central black holes in active galaxies (which traditionally equates central 
luminosities with Eddington luminosities) might have been seriously 
underestimated (C96). Fabian \& Rees (1995) raise such concerns
recently using self-similar, super-advective flow solutions. However,
since we found that the angular momentum in the accretion is
no less than $20-50$ percent of the Keplerian disk, the conclusions would be
expected to be less dramatic.

The author acknowledges the hospitality of Max-Planck Institute (Garching)
and Landessternwarte (Heidelberg) in the summer of 1994, where 
this work was partially completed. He thanks  R. Khanna and
Stefan Appl (both at Landessternwarte) for discussions. 
He also thanks an unknown referee
and Paul Wiita for the encouragement to expand a smaller, cryptic 
version of this work originally submitted to Astrophysical Journal 
Letters (also in the e-print archive under astro-ph/9508060). 
The present research is supported through a Senior Research Associateship 
award from National Academy of Sciences.

\vfil\eject

\noindent Fig. 1(a-b): Variation of specific entropy  as functions of 
specific energy $E(x_c)$ at sonic points. Location of the sonic point 
increases along the direction of the arrow on the curves. 
Solid, long-dashed and 
short-dashed curves represent saddle type, nodal type, and spiral type sonic
points respectively. Three groups of curves are drawn for the viscosity 
parameter $\alpha_P=0.0$, $0.4$ and $0.8$ (as marked) and two side by side 
curves in each group are for cooling and heating dominated (marked) cases
(for $\alpha_P=0$ these two curves coincide). $l_{in}=1.65$ and $\gamma=4/3$
are chosen. The branches of the 
type marked $AMB$ correspond to the inner sonic points, 
and branches of the type marked $CMD$ correspond to
outer sonic points. Horizontal arrows mark typical
Rankine-Hugoniot shock transitions in winds (upper) 
and accretion (lower) respectively.
In (b), the region near $M$ is zoomed to illustrate the
origin of critical viscosities. 
Sonic points slide along the curves as the viscosity and 
the cooling parameters are changed. See text for details.

\noindent Fig. 2(a-b): Variation of Mach number with logarithmic
radial distance as viscosity parameter $\alpha_\Pi$ (marked on the
upper axis) and the heating parameter $f$ (marked on the
left axis) are varied. Inner sonic point $x_{in}=2.795$ and angular momentum
constant $l_{in}=1.65$ are chosen. Note the changes in topologies
and $\alpha_\Pi$ crosses the critical viscosities.
In (b), the ratio $R$ of disk and Keplerian angular momenta are shown,
the horizontal short-dashed lines mark $R=1$. For low and high viscosities
the flow joins Keplerian disk, but for intermediate cases the disk
must have a shock or pass only though the outer sonic point to join
a Keplerian disk and the horizon.

\noindent Fig. 3(a-b): Same as Fig. 2a, but (a) location of
sonic point $x_{in}$ (marked in each box), 
and (b) angular momentum $l_{in}$ (marked on the upper axis)
and heating parameter $f$ (marked on the axis on left) are varied.

\noindent Fig. 4: Same as Fig. 2a, but the outer sonic point
is held fixed at $x_{out}=35$ while varying the viscosity and 
heating parameters.

\noindent Fig. 5(a-b): Example of a complete solution which includes a shock wave. 
$\alpha_\Pi=0.05$, $l_{\rm in}=1.6$, $\gamma=4/3$ and $f=0.5$, $x_{out}=50$ are used. 
Shock conditions are satisfied at $x=13.9$ (vertical dashed curve)
and the flow subsequently enters the black hole along the long dashed 
curves after becoming supersonic at $x_{in}=2.8695$. The arrowed 
curves trace the complete solution with a shock wave in a black hole
accretion. Double arrowed curves denote the shock solution in a neutron star
accretion with sub-sonic inner boundary condition. In (b), the 
procedure of obtaining the shock solutions is illustrated by 
plotting the Mach invariant $C$, the total pressure $\Pi$ and the 
energy $E(x)$ of the subsonic and super branches all of which
intersect at the shock location. Note that the angular momentum remains 
continuous across the shock. Other physical quantities, such as the
proton temperature and the radial velocity are also shown.

\noindent Fig. 6: Parametric dependence on the formation and
location of shocks in black hole and neutron star accretions. Single
arrows represent shocks in accretion around black holes while the 
double arrows represent shocks around compact neutron stars.
Pairs of parameters ($\alpha_\Pi, f$)
are shown in each box. The outer sonic point is chosen to be 
at $x_{out}=50$. In the neutron star case, the shock transition
takes place only to that particular sub-sonic branch which
corrotates with the star at the star surface.

\noindent Fig. 7(a-b): Ratio of (a) radial to azimuthal velocities
($v_r/v_\phi$) and (b) disk angular momentum to Keplerian angular momentum
($l/l_{Kep}$) are shown in a few
solutions. In (a) the curves marked with numbers correspond to the grid
number of Fig. 2a. The curve marked `shock' correspond to the solution
in Fig. 5a. Solutions except those marked `43' and `12' join smoothly
with Keplerian disks as they become rotation dominated. In (b), the curves
are marked with viscosity parameters (the curve marked $0.05$ correspond
to the shock solution in Fig. 5a). Note that $x_{Kep}$,
where the flow joins a Keplerian disk, depends inversely on the
viscosity parameter. See text for detail.

\noindent Fig. 8: Same as Fig. 1a, but drawn for $\gamma=5/3$ to illustrate
that the outer saddle type sonic point does not exist (a feature
shared by flows with $\gamma>1.5$ in a vertically averaged model).
Thus the flow from a Keplerian disk must pass only through the inner
sonic point. Centrifugally supported shocks could form only if the
flow at outer boundary is already supersonic (e.g., originated from
stellar winds).

\noindent Fig. 9(a-b): Similar to Figs. 2-3, showing the variation 
of solution topologies when $\gamma=5/3$ is chosen. See text for details.
 

\begin{references}

\reference Abramowicz, M.A. \& Chakrabarti, M.A. 1990, ApJ, 350, 281
\reference Abramowicz, M.A., Czerny, B., Lasota, J.P. \& Szuzkiewicz, E.
1988, ApJ, 332
\reference Abramowicz, M.A. \&  Zurek, W.H. 1981, ApJ, 246, 314
\reference Anderson, M.R., \& Lemos, J.P.S. 1986, MNRAS, 233, 489
\reference Begelman, M.C. \& Chiueh, T. 1988, ApJ, 332, 872
\reference Begelman, M.C. 1978, MNRAS, 185, 847
\reference Blandford, R.D. \& Payne, D.G. 1981, MNRAS, 194, 1033
\reference Blondin, J.M. 1986, ApJ, 308, 755
\reference Cannizzo, J.K. 1993, in Accretion Disks in Compact Stellar
Systems, ed. J. Craig Wheeler (Singapore: World Scientific).
\reference Castor, J.I., Abbott, D.C., \& Klein, R.I. 1975, ApJ, 195, 157.
\reference Chakrabarti, S.K. 1985, ApJ, 288, 1 (C85)
\reference Chakrabarti, S.K., Jin, L., \& Arnett, W.D. 1987,
ApJ, 313, 674
\reference Chakrabarti, S.K. 1989, ApJ, 347, 365 (C89)
\reference Chakrabarti, S.K. 1990a, MNRAS, 243, 610 (C90a)
\reference Chakrabarti, S.K. 1990b, {\it Theory of Transonic Astrophysical
Flows} (Singapore: World Scientific) (C90b)
\reference Chakrabarti, S.K. 1990c, ApJ, 362, 406
\reference Chakrabarti, S.K. 1990d, ApJ, 350, 275
\reference Chakrabarti, S.K. 1990e, MNRAS, 246, 134
\reference Chakrabarti, S.K. 1992, MNRAS, 256, 300
\reference Chakrabarti, S.K. in Numerical Simulations in Astrophysics, 
Eds. J. Franco, S. Lizano, L. Aguilar \& E. Daltabuit (Cambridge University
Press: Cambridge, 1993) [C93]
\reference Chakrabarti, S.K. 1994, in Proceedings of 17th 
Texas Symposium (New York Academy of Sciences, New York)  [C94]
\reference Chakrabarti, S.K. 1995, ApJ, 441, 576 [C95]
\reference Chakrabarti, S.K. 1996a, Physics Reports, 266, No. 5-6, 238 [C96a]
\reference Chakrabarti, S.K. 1996b, MNRAS, in press [C96b]
\reference Chakrabarti, S.K., \&  Molteni, D. 1993, ApJ, 417, 671 [CM93]
\reference Chakrabarti, S.K., \&  Molteni, D. 1995, MNRAS, 272, 80 
[CM95] 
\reference Chakrabarti, S.K., \& Titarchuk, L. 1995, ApJ (Dec 20th) [CT95]
\reference Chang K. M., \& Ostriker, J. P. 1985, ApJ, 288, 428  
\reference Colpi, M., Maraschi, L. \& Treves, A. 1984, ApJ, 280, 319
\reference Ebisawa, K. et al. 1994,  PASJ, 46, 375
\reference Ebisawa, K., Titarchuk, L.G., \& Chakrabarti, S.K. 1996,
 PASJ, 48, No.1 in press
\reference Englmaier, P. Diploma Thesis, University of Heidelberg (1993).
\reference Fabian, A.C., \& Rees, M.J. 1995, MNRAS, in press
\reference Flammang, R.A. 1982, MNRAS, 199,833
\reference Ferrari, A., Trussoni, E., Rosner, R. \& Tsinganos, T.
1985, ApJ, 294, 397
\reference Frank, J.,  King, A.R., \& Raine, D.J., Accretion Power 
in Astrophysics (Cambridge University Press, Cambridge, 1991)
\reference Ipser, J.R. \& Price, R.H. 1982, ApJ, 255, 654
\reference Lasota, J.P., Abramowicz, M., Chen, X., Krolik, J.H., Narayan,R., \& Yi, I. 1996, preprint
\reference Liang, E.P.T \& Thomson, K.A., 1980, ApJ, 240, 271
\reference Lu, J.F., 1985, A\&A, 148, 176
\reference Matsumoto, R., Kato, S., Fukue, J., \& Okazaki, A.T.
1984,  PASJ, 36, 71
\reference M\'esz\'aros, P. 1975, A \& A,  44, 59
\reference Meyer, F., \& Meyer-Hofmeister, E. 1983, A\&A, 128, 420
\reference Molteni, D., Lanzafame, G., \& Chakrabarti, S.K. 1994, 
ApJ, 425, 161 [MLC94]
\reference Molteni, D., Sponholz, H., \&  Chakrabarti, S.K. 1996, 
ApJ (in press) [MSC96]
\reference Muchotrzeb B. 1983, Acta Astron., 33, 79
\reference Nakayama, K.  1982, MNRAS, 259, 259
\reference Narayan, R., 1996, in ``Basic Physics of Accretion Disks" eds. S. Kato et al.,
Gordon and Breach Science Publishers (New York, 1996) in press
\reference Narayan, R., \&  Yi, I. 1994, ApJ, 428, L13 
\reference Narayan, R., \&  Yi, I. 1995a, ApJ, 444, 231
\reference Narayan, R., \&  Yi, I. 1995b, ApJ,  in press
\reference Narayan, R., Yi, I., \& Mahadevan, R. 1995, Nature, 374, 623
\reference Narayan, R., McLintock, J.E., \&  Yi, I. 1996, ApJ, in press 
\reference Nobili, L., Turolla, R. \& Zampieri, L. 1991, ApJ, 383, 250
\reference Nobuta, K. \& Hanawa, T. 1994, PASJ, 46, 257
\reference Novikov, I. \&  Thorne, K.S. in: Black Holes,
eds. C. DeWitt and B. DeWitt (Gordon and Breach, New York, 1973).
\reference Ostriker, J.P., McCray, R., Weaver, R. \& Yahil,  A. 1976,
ApJ, 208, L61
\reference Paczy\'nski, B. \& Bisnovatyi-Kogan, G. 1981, Acta Astron., 
31, 283
\reference Paczy\'nski, B. \& Wiita, P.J. 1980 A\&A, 88, 23
\reference Park, M.G., 1990, ApJ, 354, 64
\reference Park, M.G., \& Ostriker, J.P. 1989, ApJ, 347, 679
\reference Pringle, J. 1981, Ann. Rev. Astron. Astrophys., 19, 137
\reference Rees, M., M. J., Begelman, M. C., Blandford, R. D., \& 
Phinney, E. S. 1982, Nature, 295, 17
\reference Ryu, J., Brown, G., Ostriker, J. \& Loeb, A. 1995, ApJ, 
(in press). 
\reference Shakura, N.I. \& Sunyaev, R.A. 1973, A\&A, 24, 337 [SS73]
\reference Shapiro, S.L.  1973a, ApJ, 180, 531
\reference Shapiro, S.L.  1973b, ApJ, 1185, 69
\reference Shapiro, S.L. \& Teukolsky, S.A., Black Holes, White Dwarfs 
and Neutron Stars --- the Physics of Compact Objects (John Wiley \& Sons, 
New York, 1983)
\reference Sunyaev, R. A. et al., 1994, Astron. Lett. 20, 777
\reference Spruit, H.C. 1987, A \& A, 184, 173
\reference Taam, R.E. \& Fryxell, B.A., 1985, ApJ, 294, 303
\reference Takahashi, M., Nitta, M., Tatematsu, Y., Tomimatsu, A.
 1990, ApJ, 363, 206
\reference Tanaka, Y. in Proceedings of the 23rd ESLAB symposium, ed. 
J. Hunt \& B. Battrick, vol. 1, p.3, 1989, (ESA, Paris)
\reference Thompson, J.M.T. \&  Stewart, H.B., Nonlinear Dynamics and Chaos
(John Wiley \& Sons Ltd., 1985)
\reference Wandel, A., Yahil, A. \& Milgrom, M., 1984, ApJ, 282, 53

\end{references}
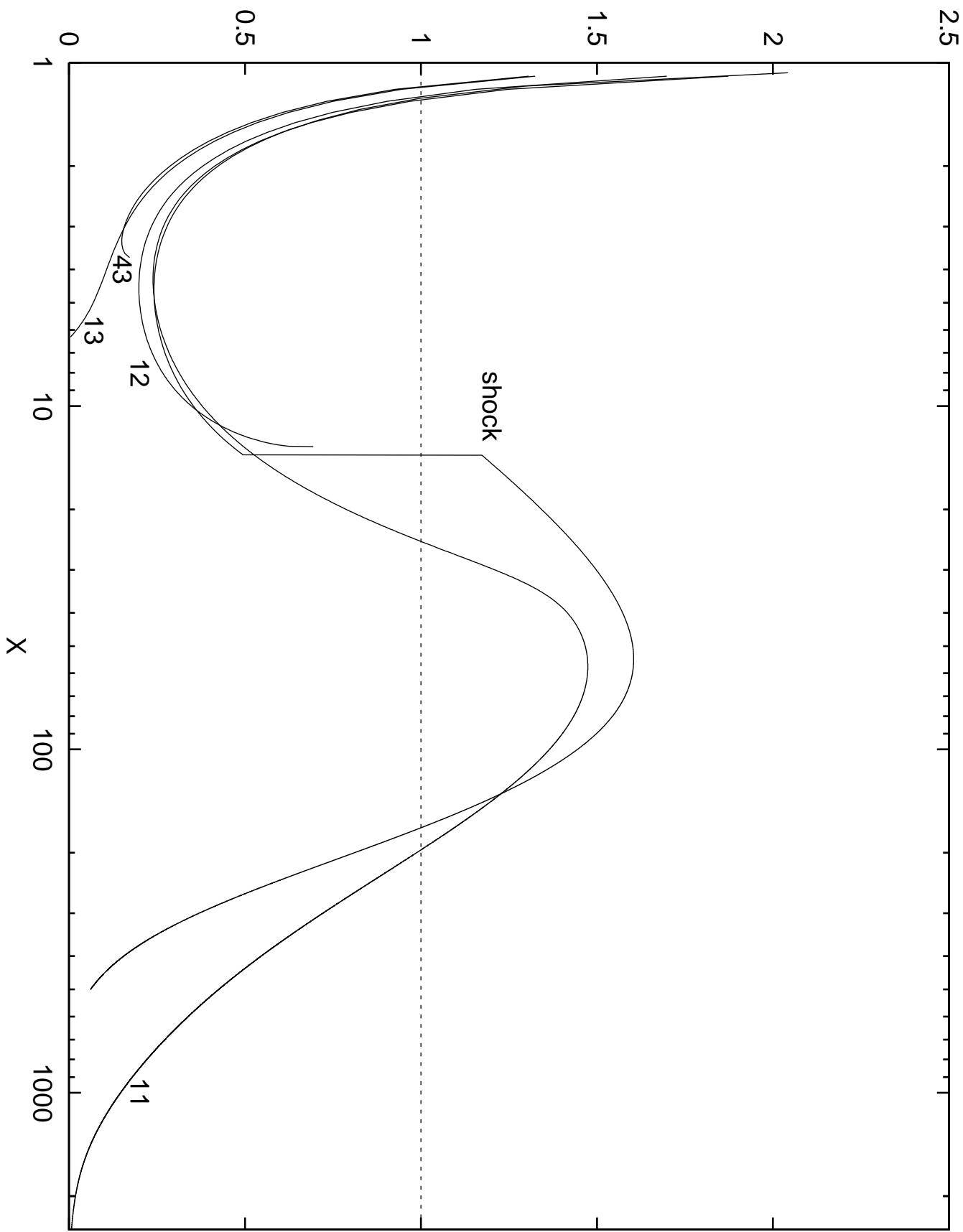
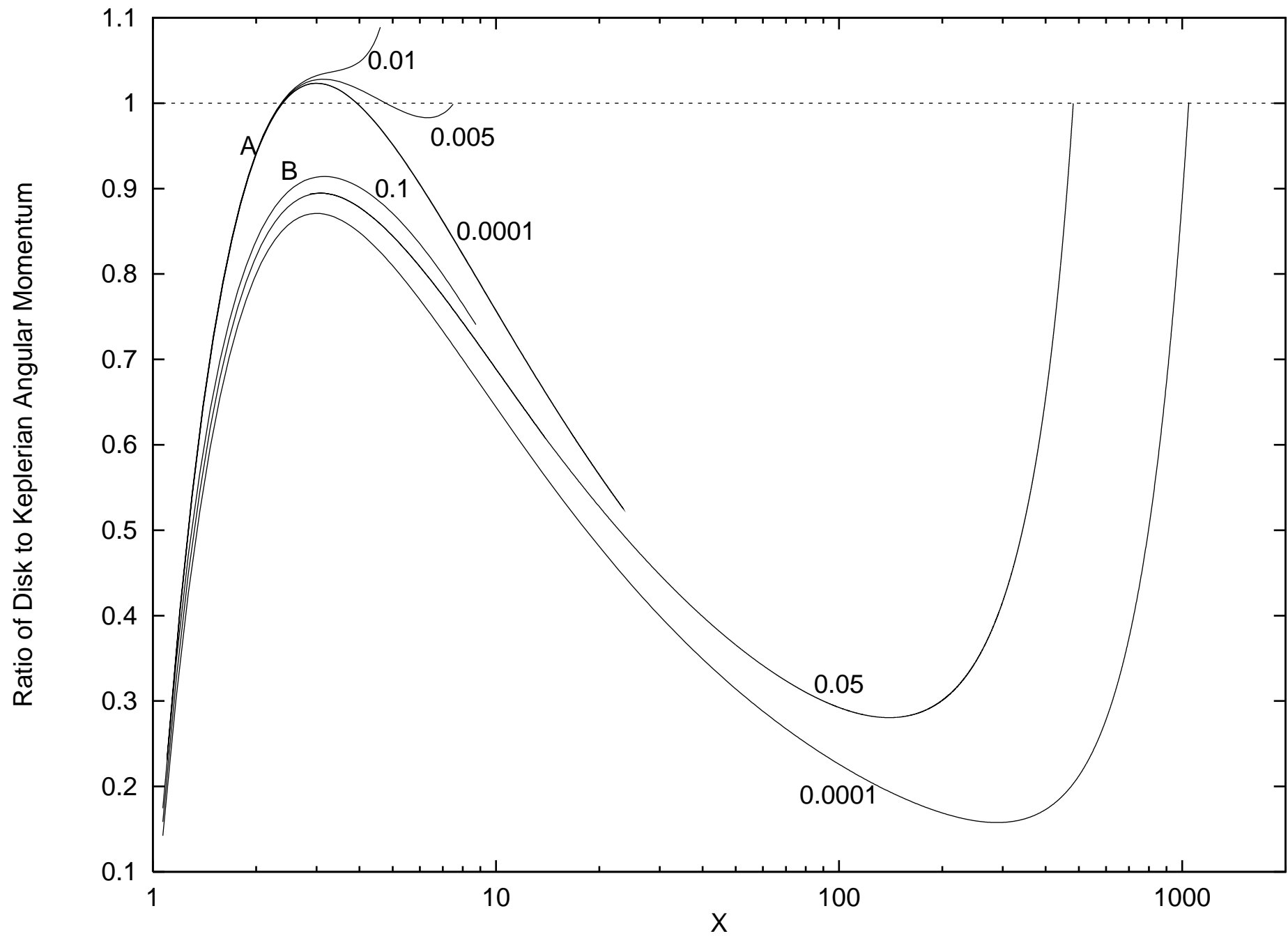
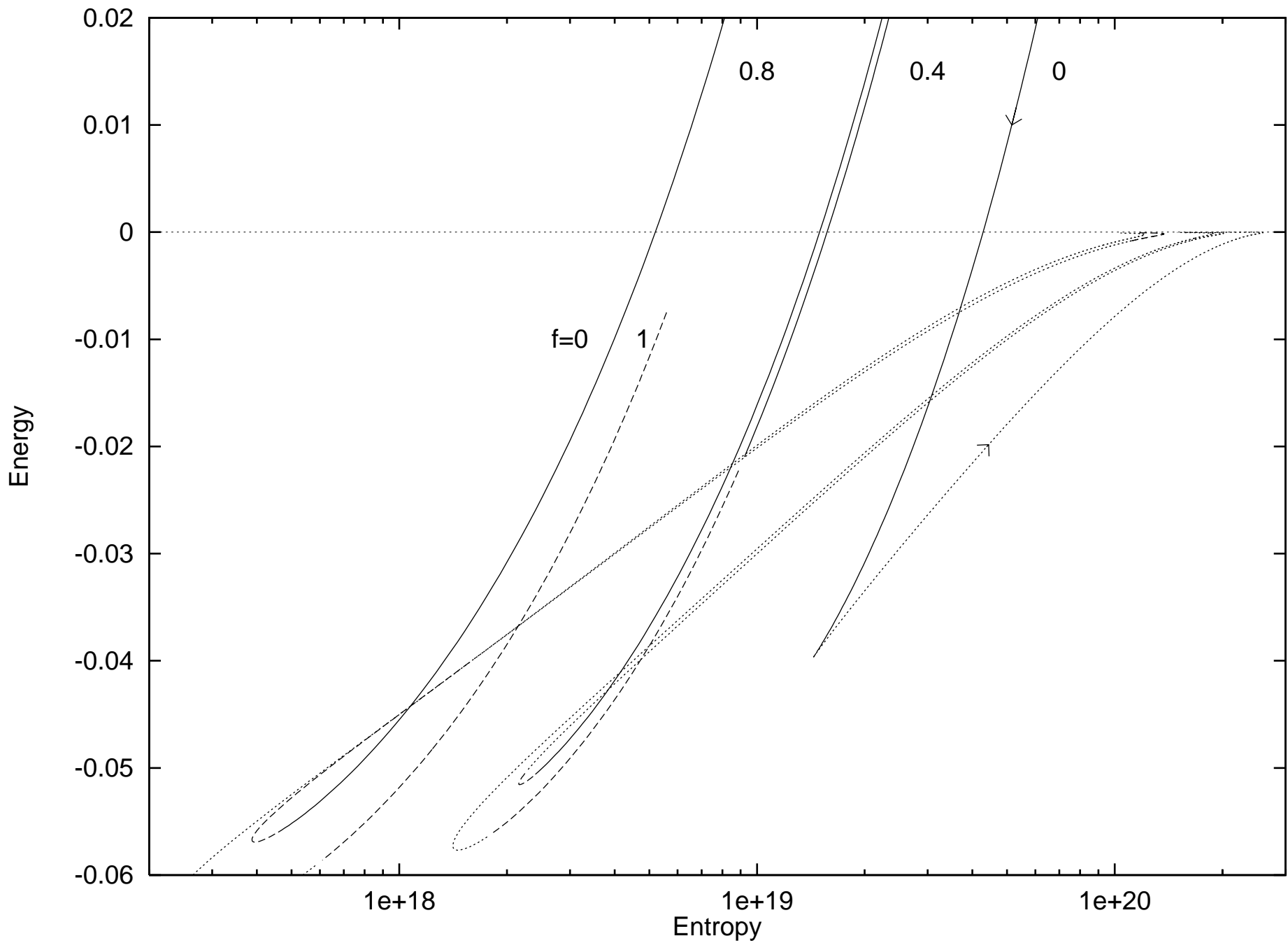
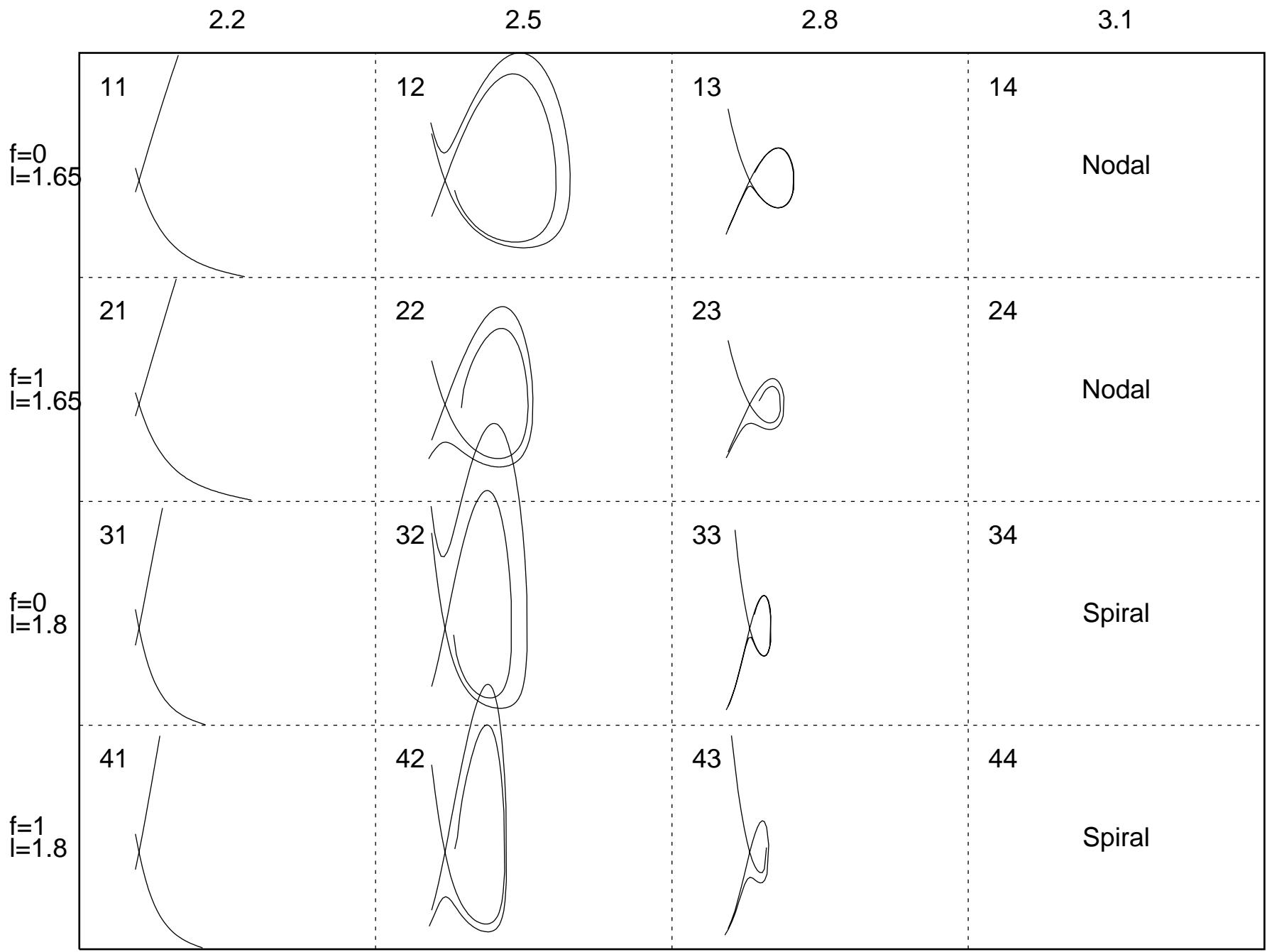
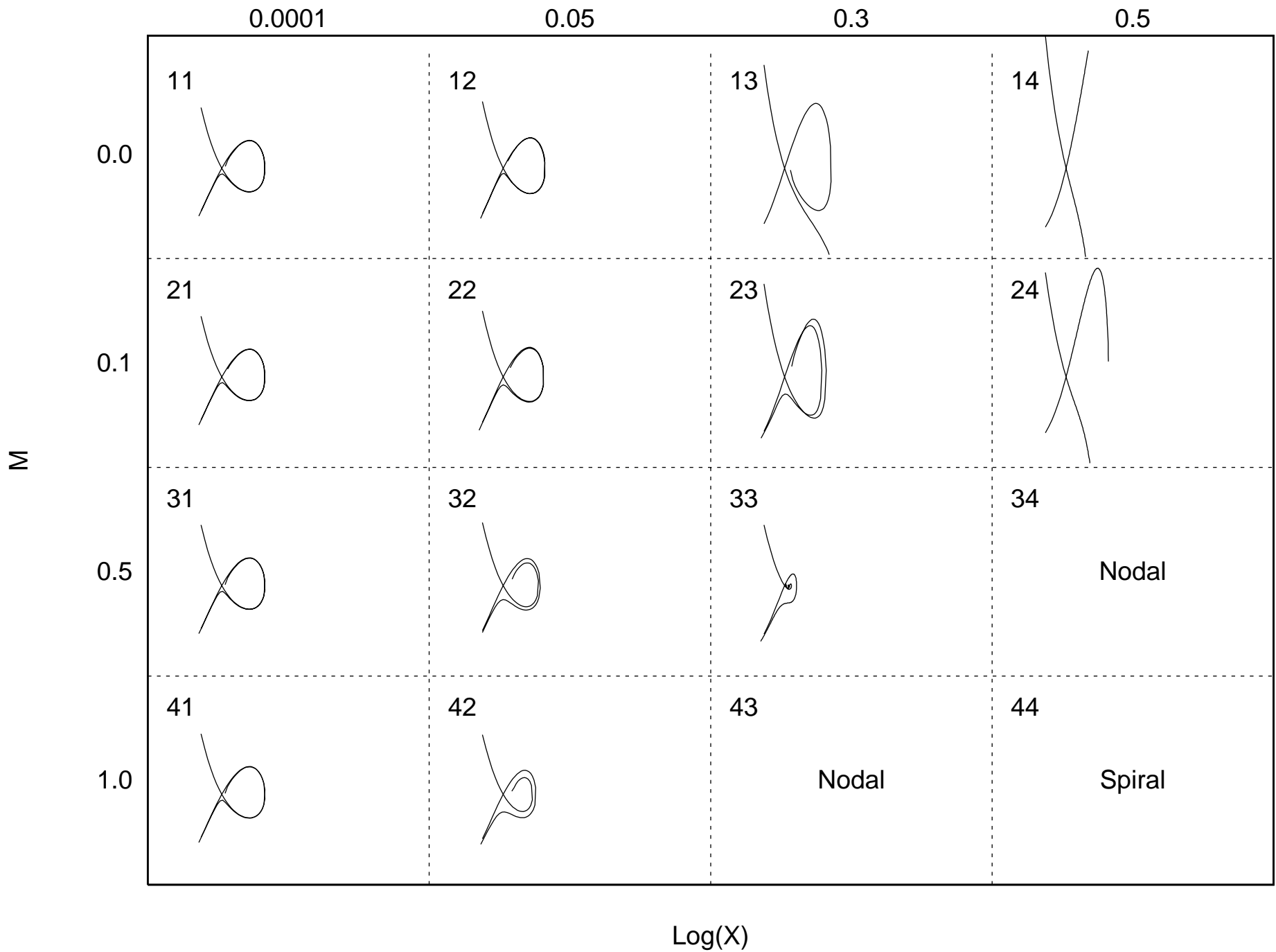
\end{document}